\documentclass[11pt,3p]{elsarticle} 

\usepackage{hyperref}
\usepackage{booktabs}
\usepackage{calc}
\usepackage{graphicx}
\usepackage{subcaption}
\usepackage{multirow}
\usepackage{algpseudocode,algorithm,algorithmicx}
\usepackage{comment}
\usepackage{paralist}
\usepackage{color}
\usepackage{xcolor}
\usepackage{amsmath}
\usepackage{amssymb}
\usepackage{amsthm}
\usepackage{xspace}
\usepackage[inline]{enumitem}
\usepackage{mathtools}
\usepackage{bm}
\usepackage{lipsum}
\usepackage{lineno}
\usepackage{hyperref}
\usepackage{setspace}
\usepackage{yhmath}
\usepackage[labelformat=simple]{subcaption}

\graphicspath{{figures/}}

\newcommand*{\unit}[1]{\ensuremath{\mathrm{\,#1}}}

\newcommand{\dw}{DecaWave\xspace}
\newcommand{\iw}{\ensuremath{I_w}\xspace}
\newcommand{\e}{endpoint\xspace}
\newcommand{\es}{endpoints\xspace}
\newcommand{\hnh}{HnH\xspace}
\newcommand{\ma}{MA\xspace}
\newcommand{\mas}{MAs\xspace}
\newcommand{\sa}{SA\xspace}
\newcommand{\sas}{SAs\xspace}
\newcommand{\gd}{GD\xspace}
\newcommand{\gds}{GDs\xspace}
\newcommand{\iwd}{\ensuremath{\frac{\iw}{2}}\xspace}

\newcommand{\drf}{\textsc{Drf}\xspace}
\newcommand{\drfe}{\textsc{DrfE}\xspace}

\newcommand{\lee}{\textsc{Lee}\xspace}
\newcommand{\xiao}{\textsc{Xiao}\xspace}

\newcommand{\st}{\ensuremath{\Pi}\xspace}
\newcommand{\home}{{\sc{Home}}\xspace}

\newcommand{\rmin}{\ensuremath{r\textsubscript{min}}\xspace}
\newcommand{\amin}{\ensuremath{\alpha\textsubscript{min}}\xspace}

\newcommand{\revision}[1]{{\color{black}{#1}}}

\journal{Journal of Pervasive and Mobile Computing}

\begin{document}

\begin{frontmatter}
	
\title{A Comprehensive Investigation on Range-free Localization Algorithms \\ with Mobile Anchors at Different Altitudes 
}


\author[address_mst]{Francesco Betti Sorbelli}
%
\author[address_mst]{Sajal K.~Das}
%
\author[address_unipg]{Cristina M.~Pinotti}
%
\author[address_unifi]{Giulio Rigoni}
%
\address[address_mst]{Dept. of Computer Science, Missouri University of Science and Technology, Rolla, MO, USA}
\address[address_unipg]{Dept. of Computer Science and Math., University of Perugia, Italy}
\address[address_unifi]{Dept. of Computer Science and Math., University of Florence, Italy}
\begin{abstract}
In this work, the problem of localizing ground devices (GDs) is \revision{studied} comparing the performance of four
range-free (RF) localization algorithms that use a mobile anchor (MA).
All the investigated algorithms are based on the so-called heard/not-heard (HnH) method,
which allows the GDs to detect the MA at the border of their antenna communication radius.
Despite the simplicity of this method, its efficacy in terms of accuracy is poor because it relies on the
antenna radius that continuously varies under different conditions.
Usually, the antenna radius declared by the manufacturer
does not fully characterize the actual antenna radiation pattern.
In this paper,
the radiation pattern of the commercial DecaWave DWM1001 Ultra-Wide-Band (UWB) antennas is observed in a real test-bed at different altitudes for collecting more information and insights on the antenna radius.
The compared algorithms are then tested using both the observed and the manufacturer radii.
The experimental accuracy is close to the expected theoretical one only when the antenna pattern is actually omnidirectional.
However, typical antennas have strong pattern irregularities 
that \revision{decrease} the accuracy.
For improving the performance,
we propose range-based (RB) variants of the compared algorithms in which,
instead of using the observed or the manufacturer radii,
the actual measured distances between the \ma and the \gd are used.
The localization accuracy tremendously improves confirming that
the knowledge of the exact antenna pattern is essential for any RF algorithm. 

\end{abstract}

\begin{keyword}
	Drone \sep 
	rover \sep
	localization \sep
	range-free \sep 
	range-based \sep
	antenna model
\end{keyword}

\end{frontmatter}

\section{Introduction} 
Considering the \revision{recent} interest in the Internet of Things (IoT), and in general in the Wireless Sensor Networks (WSNs), the problem of localizing sparse wireless sensors is more and more appealing to researchers. 
Due to their limited size and low cost, those sensors can be installed everywhere for any specific task. 
For instance, localization can help to search and rescue people after natural disasters, to monitor the structural health of building, or even to track people during the COVID-19 pandemic in order to decrease the circulation of the virus. 

Technically speaking, the localization problem aims at estimating the position of {\em ground devices} (\gds) deployed on the WSN. 
For this important task, special devices whose positions are known a-priori, called \emph{anchors}, are in charge of localizing the \gds. 
In the literature, localization can be classified in respect to the type of anchors deployed, i.e., \emph{static anchors} (\sas) or \emph{mobile anchors} (\mas). 
The former scenario requires a discrete amount of \sas with fixed positions, while the latter one requires a single \ma (e.g., rover, drone) with GPS capabilities (need of fresh coordinates while moving). 
Notice that the usage of \sas is not affordable in terms of costs when the WSN is large; analogously, such a fixed infrastructure cannot be quickly reused elsewhere. 
Hence, in this paper we consider only the localization with a single \ma.
Localization algorithms can be also broadly categorized as \emph{range-free} (RF) or \emph{range-based} (RB) approaches~\cite{han2016survey}. In the previous, the position is estimated only by discovering if the \gd and \ma are one in the range of the other. In the latter, the position of the \gd is estimated by taking measurements (e.g., distance, angle, signal strength) between it and the \ma. 
Usually, RB algorithms are known to be more accurate than RF ones but at the cost of additional specialized hardware. Finally, among the RF algorithms, the {\em radius-based} algorithms assume the knowledge of the transmission radius, while the {\em radius-free} ones do not.

In this paper, we evaluate on a test-bed the accuracy \revision{of many RF algorithms}.
They exploit the {\em heard/not-heard} (\hnh) method, consisting of detecting two consecutive messages transmitted by the \ma, one heard and one not-heard from the \gd, called {\em \es}.
All the algorithms, but one which is new, have been described in the literature under the {\em ideal model}. 
The ideal model assumes that both \ma and \gd are equipped with an isotropic antenna whose radiation pattern is a perfect sphere. 
Consequently, the transmission and receiving areas, which are given by the intersection of the sphere with the earth surface, are perfect circles. 
The ideal model can be emulated in reality only with expensive hardware on noiseless areas, which are not the usual operating conditions of most localization applications that occur, for example, in precision agriculture or search-and-rescue context.
To reproduce the typical operating conditions, we adopt for our test-bed a set of inexpensive \dw DWM1001 Ultra-Wide-Band (UWB) antennas and an off-the-shelf 3DR Solo drone as the flying \ma.
%
Our pursued goals can be \revision{summarized} as follows:
\begin{itemize}
    \item \revision{We experimentally evaluate the performance of many localization algorithms}, with very little knowledge on the antenna radiation pattern; 
    \item We study how the \ma's altitude can affect the localization performance. 
\end{itemize}
The experimental results raised difficulties, doubts, and questions.
Firstly, collecting a large set of \es on the field has proven to be challenging. 
So, we search for the \revision{statistical} distribution that best fits the already observed \es for generating a large new set of synthetic \es. \revision{Such a set is used for testing the compared algorithms.}
To mitigate the poor observed performance in the real scenario,
we decide to exploit the DWM1001's capability of \revision{taking} distance measurements to gain awareness of the antenna radiation pattern. 
\revision{Despite this strategy makes the RF algorithms actually RB,}
the original algorithmic rules are kept, and the awareness of the exact antenna radiation pattern, though incomplete, significantly improves the performance.

   
 


	

The paper, which is an extension on~\cite{sorbelli2020range}, is structured as follows: 
Sec.~\ref{sec:related} reviews the literature on localization.
Sec.~\ref{sec:algorithms} presents the compared algorithms.
Sec.~\ref{sec:exp-setup} describes the test-bed.
Sec.~\ref{sec:antenna-analysis} analyzes the antenna's performance.
Sec.~\ref{sec:simulation} evaluates the algorithms under different scenarios.
Sec.~\ref{sec:range-based} shows how the algorithms can incorporate the distance measurements. 
Sec.~\ref{sec:conclusion} offers conclusions.

\section{Related Work}\label{sec:related}
In this section we cover the state-of-the-art about localization in WSNs, giving first a brief overview about known techniques with \sas, \revision{and then surveying more in detail} techniques using a \ma. Finally, we describe a few works that implemented real test-beds about localization.

\paragraph{The Static Anchors Scenario}\label{sec:static-anchor-scenario}
In literature, different algorithms exist to tackle the issue of localization of \gds using \sas. DV-HOP based techniques~\cite{xiao2017rssi} and Amorphous algorithms~\cite{zhao2014amorphous} approximate the distance 
between \gds using the number of hops between them and estimating the average hop distance inside the WSN. Then, using the trilateration method, each \gd computes its own position.
The Centroid algorithm used in~\cite{wang2011weighted} is another technique where each \sa broadcasts its position to the surrounding, and each \gd computes its own position calculating the average of all the coordinates of  the \sas that it can hear. 
Other works propose solutions based on the Approximate Point in Triangulation (APIT)~\cite{zeng2009improvement} whose goal is to divide the area in triangles in which unknown \gds reside. In each overlapped section  (i.e., a polygon) of triangle area, the position of the unknown \gd is computed calculating the centroid. 
The main issue for these solutions is the relative high number of \sas required for an acceptable localization error.

\paragraph{The Mobile Anchor Scenario}\label{sec:mobile-anchor-scenario}
All these algorithms emulate multiple \sas with a single \ma that continuously broadcasts its position.
For localizing \gds, a \ma has to plan a route (static path) in advance inside the WSN. On that route, the \ma estimates the distance between itself and the \gds in range, and eventually the \gds' positions are computed performing trilateration.
In~\cite{koutsonikolas2007path} three different 2D movement trajectories have been studied, i.e., SCAN, DOUBLE-SCAN, and HILBERT.
The distance between two consecutive segments of the trajectories is defined as the resolution.
The simplest algorithm is SCAN, 
in which the \ma follows a path formed by vertical straight lines interconnected by horizontal lines.
Essentially, the \ma sweeps the area along the $y$-axis.
The main drawback is that it provides a large amount of collinear \es.
In order to resolve the collinearity problem, DOUBLE-SCAN sweeps the 
sensing area along both the $x$-axis and the $y$-axis.
However, in this way the path length is doubled compared with SCAN.
Finally, a level-$n$ HILBERT curve divides 
the area into $4^n$ square cells and connects the centers of those
cells using $4^n$ line segments, each of length equal to the length
of the side of a square cell.
Generally, HILBERT provides more non-collinear \es but the path length can be very long if the resolution increases.
All the above techniques are based on straight lines and suffer \revision{from} collinearity problem.
In order to heavily reduce collinearity, S-CURVES~\cite{huang2007static} has been introduced,
which is similar to SCAN except that it uses curves rather than straight lines.
Even though the collinearity problem is almost resolved, the main problem is that it does not properly cover the four corners of the squared sensing area.
\revision{One of the best techniques is LMAT~\cite{jiang2011lmat}.}
The main idea is to plan a series of \mas in such a way as to form equilateral triangles, avoiding collinearity problems.
Each \gd inside a triangle is localized by a trilateration procedure using the three vertices.


All the previous algorithms \revision{are range-based since they rely on measurements (signal strength, distance, etc.)}. \revision{Only a few range-free} techniques have been proposed \revision{in the literature}. Among them, the ones proposed by Xiao~\cite{xiao2008distributed} and Lee~\cite{lee2009localization} rely on a ground \ma, while the one proposed by Betti~\cite{bettisorbelli2019rangefree}, relies on a flying \ma. We will give more details and insights about those algorithms in Sec.~\ref{sec:algorithms}.

\paragraph{Implemented Test-beds}
\revision{To} the best of our knowledge, only a few test-bed implementations have been done aimed at comparing the performance among RF algorithms with a \ma.
Recently in~\cite{bettisorbelli2019ground}, a test-bed using inexpensive UWB antennas and a drone as \ma
evaluates the localization accuracy of 
the RF \drf algorithm proposed in~\cite{bettisorbelli2019rangefree}.
Such an algorithm strictly relies on the good quality of the antenna radiation pattern and requires
\revision{simple} geometrical rules for estimating the \gd's position.
Unfortunately, in practice, the experimental localization error 
obtained by the implemented \drf algorithm is large.
%
Another recent study in~\cite{chen2018impact} shows that 
the irregularities of the hardware antenna radiation pattern can heavily 
affect the air-to-ground (A2G) link quality
between a \ma (drone) and a \gd. 
The authors study the A2G link quality of  BroadSpec UWB antennas from Time Domain Inc
by observing the Received Signal Strength Indication (RSSI).
They show a dependency \revision{of} the link quality \revision{on} the antennas' orientations, 
their elevation, and their distance.
The authors in~\cite{khawaja2019ultra} report the UWB A2G propagation channel measurements 
in an open field using a drone.
Three scenarios \revision{with different obstacles are considered} while a drone was orbiting above a \gd
at different altitudes and ground distances.
Also, different antenna orientations are considered for the drone.
Experimental results show that the received power is highly dependent on the
antenna gain of the line-of-sight (LoS) component in the elevation plane when the 
antennas are aligned (same orientation).
In a work proposed in~\cite{sinha2019impact}, authors investigate a time difference of arrival (TDoA)
based approach for localizing drones taking into account 
a simple A2G 3D antenna radiation pattern.
Experimental results show that accounting for antenna effects
makes a significant difference and reveals many important
relationships between the localization accuracy and the
altitude of the drone.
Most importantly, they finally show that the localization
performance varies in a non-monotonic pattern with
respect to the drone altitude.

\paragraph{Motivations}
Considering that the RF algorithms have been around for a long time,
that the algorithm tested in~\cite{bettisorbelli2019ground} on the field \revision{using} a drone (i.e., flying \ma)
\revision{resulted in} bad performance,
and that the irregularities showed in~\cite{chen2018impact, khawaja2019ultra, sinha2019impact} are for A2G links,
we \revision{started to think} that altitude could be one of the main cause of poor results
for the algorithm tested in~\cite{bettisorbelli2019ground} and in general for RF algorithms.
Therefore, these considerations motivate us to investigate more 
concerning the accuracy of RF algorithms at different altitudes and the quality of antennas.

\section{The Range-free Algorithms}\label{sec:algorithms}
In this section, we describe the RF algorithms that we compare. We start defining a {\em rover} as a \ma that moves on the ground, and a {\em drone} as a \ma that flies \revision{in} the sky.
During the localization procedure, called {\em mission}, the \ma visits specific points in its path, called {\em waypoints}.
Along this path, the \ma continuously transmits a {\em beacon} that can be used by any \gd within the communication range for estimating its position.
The beacon includes the current \ma's GPS position and it is sent at regular intervals of time. 
The distance among any two consecutive transmitted beacons is called {\em inter-waypoint} distance \iw and depends on the \ma's current speed.
All the algorithms that we study exploit the \hnh method relying on the detection of \es.
Performing \hnh, the \gd learns that the \ma is currently transmitting at its transmission area border.
Moreover, the \gd learns that the last (first) not-heard beacon is at a distance \iw from the first (last) heard beacon (\e), and hence at a distance no more \iw from the edge of the transmission area.
In general, three applications of \hnh are required for localizing the \gd.

In the following, we describe three RF algorithms that we evaluate, namely, \drf~\cite{bettisorbelli2019rangefree}, \xiao~\cite{xiao2008distributed}, and \lee~\cite{lee2009localization}.
We also introduce a new variant of \drf, called \drfe.
In these algorithms, the \ma travels along a static path for localizing the \gd.
The static path is formed by a sequence of consecutive straight segments in which the \ma regularly broadcasts message beacons (its current position) that the \gd is listening to.
Once the \gd has collected enough information, it can finally compute and estimate its position according to the performed algorithm.

\begin{figure}[htbp]
	\centering
	\small
	\subfloat[\drf.]{%
		\def\svgscale{0.5}
		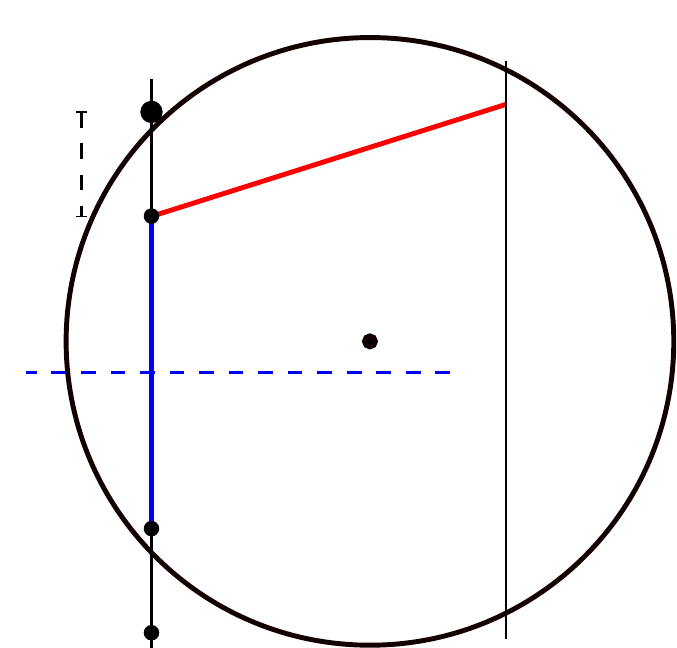
		\label{fig:model-drf}
	}
	\subfloat[\xiao.]{%
		\def\svgscale{0.5}
		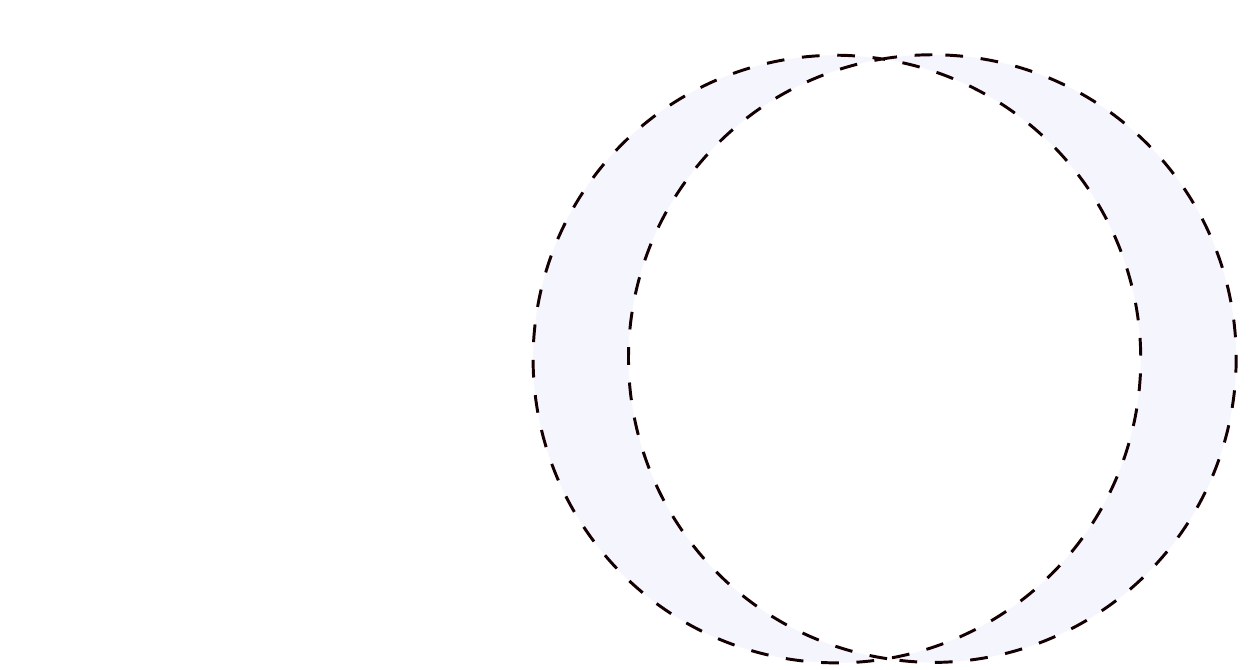
		\label{fig:model-xiao}
	}
	\subfloat[\lee.]{%
		\def\svgscale{0.5}
		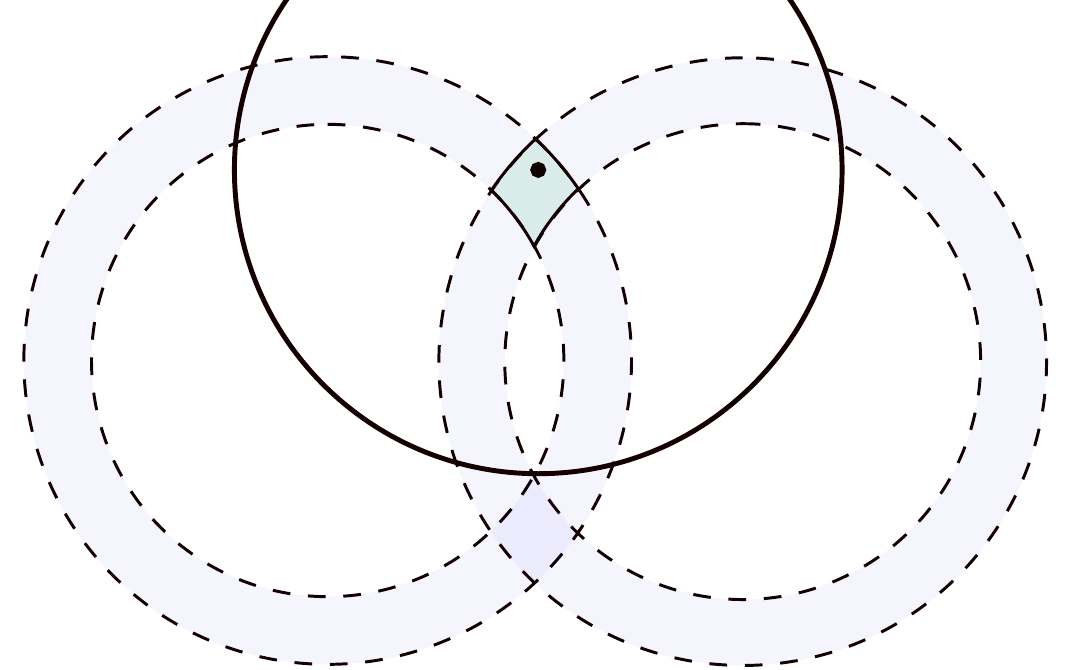
		\label{fig:model-lee}
	}
	\caption{The \drf, \xiao, and \lee localization algorithms. 
		In \xiao e \lee there are two symmetric intersection areas: a third point (not illustrated) 
		is required to find and disambiguate the intersection area where \gd resides.}
	\label{fig:intersection_areas}
\end{figure}

\subsection{The \drf Algorithm}
\drf~\cite{bettisorbelli2019rangefree} is a lightweight RF radius-free algorithm designed for drones.
This algorithm is based on the notion of {\em chord}.
In geometry, the perpendicular bisector of any circle's chord passes through the center $O$ of the circle itself.
So, the bisector of another non-parallel chord and the previous one intersect at $O$ point.
In Fig.~\ref{fig:model-drf}, the \gd is located at the point $O$, while initially the \ma travels along the segment that intersects the points in sequence $A_0$ first and \revision{then} $A_3$.
The radio receiving area of the \gd is identified by the circle centered at $O$, so if the \ma transmits a message outside that circle (e.g., in $A_0$), the \gd cannot hear any message.
However, when the \ma crosses such a circle and transmits in $A_1$, the \gd can now receive and record messages, because the relative distance between them is less than or equal to the transmitting/receiving radius.
The same reasoning can be applied for the points $A_2$ (heard) and $A_3$ (not heard).
Accordingly, the first chord is denoted by the segment with \es $A_1A_2$.
It is easy to understand that when the \ma crosses the \gd's receiving area along another segment (e.g., the one that intersects the point $B_1$), another \e is detected, and eventually two chords are identified by the pairs $A_1A_2$ and $A_2B_1$.
Finally, the \gd starts to estimate its position once it has detected these two chords computing first the associated perpendicular bisectors (dashed lines) and then their intersection point (which can be different from point $O$).
The detection of chords incurs several problems that eventually affect the localization accuracy.
Recalling that the \ma regularly broadcasts its current position (waypoint) at discrete intervals of time and that two consecutive waypoints are at distance \iw, the endpoints of the chords may not exactly fall on the circumference of the receiving disk, even if the receiving disk is a perfect circle (e.g., $A_2$ and $A_3$).

\subsection{The \xiao Algorithm}
\xiao~\cite{xiao2008distributed}
is a RF radius-based localization algorithm initially developed for ground \mas.
Like \drf, the \xiao algorithm exploits the \hnh method in order to detect special points used for building a constrained area that bounds the \gd's position.
Unlike \drf, \xiao also uses the value of the communication radius $r$.
In Fig.~\ref{fig:model-xiao}, the \gd is located at the point $O$ while the \ma travels along the segment that intersects first the point $A_0$ and then $A_3$.
Once applied the \hnh method, the \gd initially detects the first pair of heard endpoints $A_1$ and $A_2$.
However, since the segment lies on a straight line and the value of \iw is known, the \gd can also compute two additional non-heard beacons, i.e., $A_0$ and $A_3$, associated with $A_1$ and $A_2$, called pre-arrival and post-departure.
Then, four circles of radius $r$ centered at each of these four points are drawn.
Those circles create two symmetrical intersection areas (e.g., the first one is bounded by the points $P_1,P_4,P_2,P_3$) where the \gd may reside.
Hence, the \gd's position can be at the ``center'' of one of the two intersection areas.
In order to disambiguate in which intersection area the \gd resides, a third \hnh beacon is required, and the final estimated position is the one which has the closest distance, from that third point, to the radius.
The definition of center varies depending on whether the intersection area is delimited by four or five vertices~\cite{xiao2008distributed}.

\subsection{The \lee Algorithm}
\lee~\cite{lee2009localization}
is a RF radius-based algorithm very similar to \xiao.
It builds a  constrained area using the \hnh method and the knowledge of both $r$ and \iw, similarly to \xiao.
In Fig.~\ref{fig:model-lee}, once the \gd has detected the two extreme endpoints ($A_1$ and $A_2$), it traces two circles of radius $r$ and $r-\iw$ on both the points, which create two annuli, intersecting in two distinct and symmetrical intersection areas (e.g., one is bounded by the points $P_1,P_4,P_2,P_3$).
Finally, \gd resides at the center of one of such areas, and a third \e is used to disambiguate the correct one.

\subsection{The Proposed \drfe Algorithm}
Now we present a new RF algorithm, called \drfe,  
that shares the ``chord'' idea with the  \drf algorithm
and the radius information $r$ with \xiao and \lee.
In the special case that the two \es $A_1$ and $A_2$ that delimit the chord 
exactly lie on the circumference,
the \gd resides on the point $O$ of the perpendicular bisector 
that is distant $r$ from the two \es.
Precisely, there are two \revision{possible locations for} $O$, one on the left and one on the right of the chord.
Usually, using a third \e non-collinear with $A_1$ and $A_2$, 
it is possible to disambiguate the correct intersection. 

\begin{figure}[htbp]
	\centering
	\small
	\def\svgscale{0.5}
	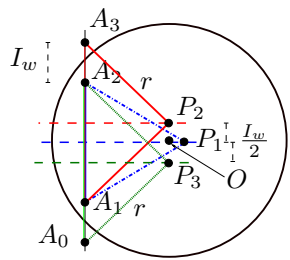
	\caption{The points $P_1$, $P_2$, and $P_3$ of the \drfe algorithm.}
	\label{fig:model-drf_evo}
\end{figure}

In general, since the \ma's path is sampled with discrete beacons at distance \iw among them, $A_1$ and $A_2$ may not \revision{lie} on the circumference and thus \gd may not exactly reside on the perpendicular bisector of the chord $\overline{A_1A_2}$,
but in its vicinity.
So, to find the \gd's position, \drfe  repeats the above construction for the three chords $\overline{A_1A_2}$,
$\overline{A_0A_2}$, and $\overline{A_1A_3}$, where $A_0$ and $A_3$ are the two non-heard beacons associated 
with the \es $A_1$ and $A_2$,
as illustrated in Fig.~\ref{fig:model-drf_evo}.
From the three chords $\overline{A_1A_2}$, $\overline{A_1A_3}$, and $\overline{A_0A_2}$, 
three intersection points $P_1$, $P_2$, and $P_3$ are obtained
at distance $r$ from the \es of their chord.
In other words, $P_1$, $P_2$, and $P_3$
form three isosceles triangles $\triangle({A_1A_2P_1})$,
$\triangle({A_1A_3P_2})$, and $\triangle({A_0A_2P_3})$
with  two oblique sides of equal length $r$. 
As before, this construction finds three vertices on the left of the chord and three vertices 
on the right that will be disambiguated using another non-collinear \e (not illustrated). 
Let us suppose that \gd is on the right of the chord  $\overline{A_1A_2}$.
\drfe places \gd at the centroid of $P_1$, $P_2$, and $P_3$ on the right of the chord  $\overline{A_1A_2}$.
Since the RF radius-based algorithms require the knowledge of the transmission radius,
in the next next we describe how to collect such information in a real test-bed.

\section{Test-bed Setup}\label{sec:exp-setup}
In this section, we describe the test-bed and the used hardware.
We fix a Cartesian coordinate system with origin at the special position \home  $(0, 0, h_0)$,
with $h_0 = 1 \unit{m}$.
At \home, we place the \gd's antenna at the top of a tripod of height $h_0$.
Also, the \ma is equipped \revision{with} an antenna.
When we set $h=0 \unit{m}$, 
we refer to a rover mission where the \ma and the \gd are placed at $h_0$,
while when we set $h>0 \unit{m}$,
we refer to a drone mission flown at an altitude $h_0+h$.

\begin{figure}[htbp]
	\centering
	\subfloat[The drone and the ground device $P$.]{%
		\centering
		\def\svgscale{0.6}
		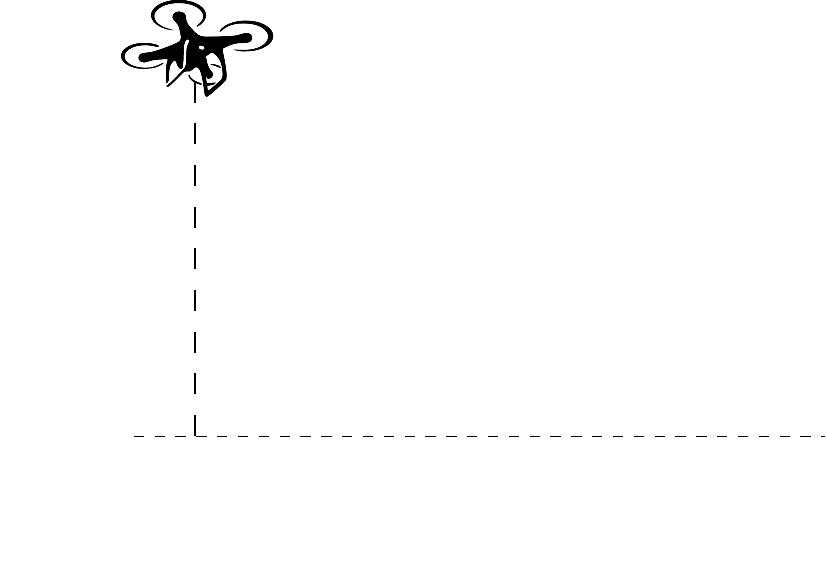
		\label{fig:ranging_precision}
	}
	\subfloat[The \ma's random path.]{%
		\small
		\centering
		\def\svgscale{0.55}
		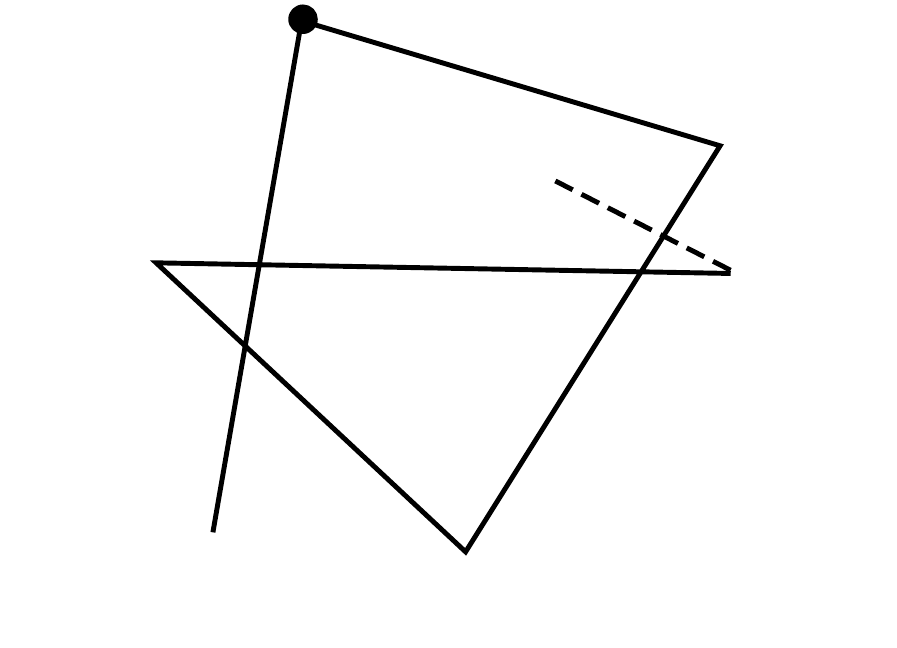
		\label{fig:path_random}
	}
	\subfloat[VV and VH.]{%
		\centering
		\def\svgscale{0.4}
		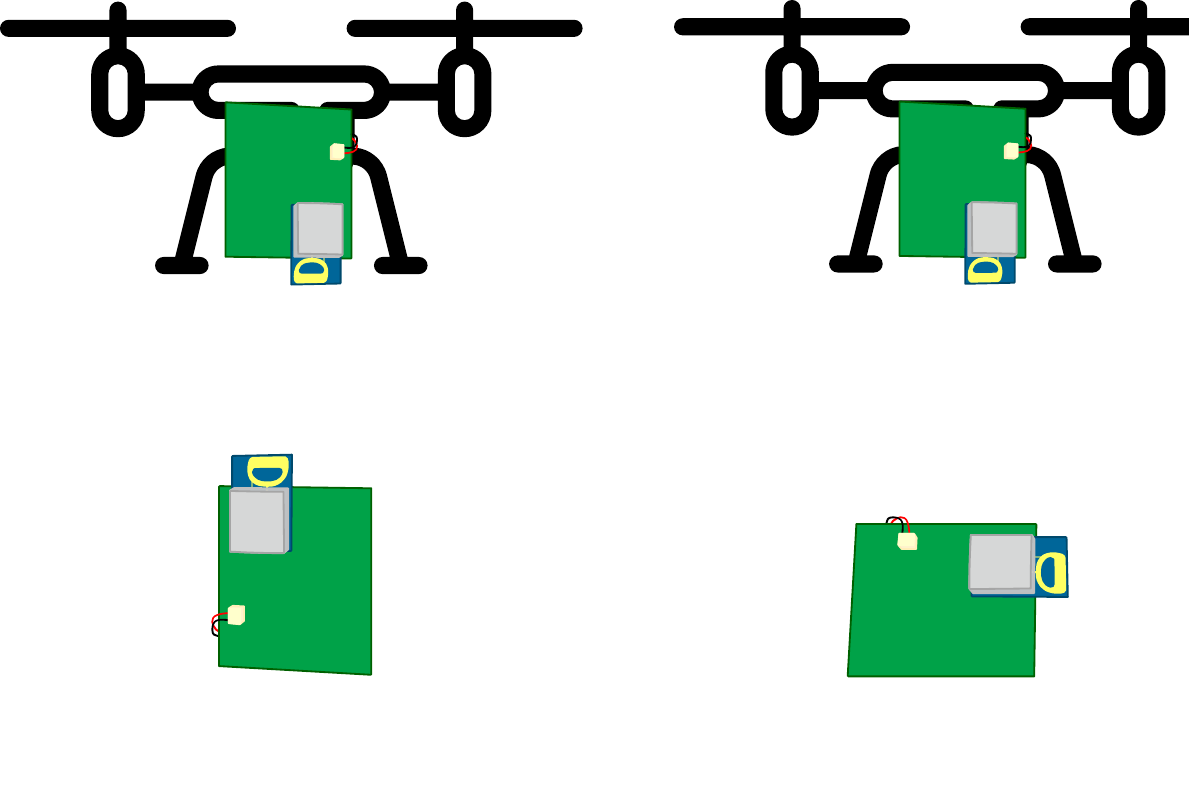
		\label{fig:setting}
	}
	\caption{The test-bed setup.}
	\label{fig:setup}
\end{figure}

In our test-bed, we employ different hardware components:
a few \dw DWM1001 UWB antennas~\cite{www-decawave}, a Raspberry Pi, and a 3DR Solo drone~\cite{www-3dr}.
According to \dw, the transmission radius is $60 \unit{m}$ and hence, in this paper, we set the manufacturer radius to $r_0 = 60 \unit{m}$.
The main component that pilots the \ma and sends commands to the \gd is the Raspberry Pi.
The Raspberry can be used for the experiments on the ground using a rover, 
or together with the drone for the aerial ones.
In the former case, i.e., \ma as a rover, we simulate the rover's behavior \revision{by} just walking in the field at a regular walking speed (about $3 \unit{km/h}$ as measured by a smart-watch)
keeping the Raspberry on the hands at $h_0 = 1 \unit{m}$ above the ground.
Moreover, since the rover has to send GPS positions, 
we rely on a cheap USB GPS module connected to the Raspberry.
Finally, in the latter case, i.e., \ma as a drone, 
we use the 3DR Solo drone
which is able to fly up to $25 \unit{min}$~\cite{www-3dr}.

A mission is a static path \st that consists of $n$ segments $S_i$, $i = 0, \ldots, n-1$.
Such a list of $n$ segments is made by generating $n+1$ random points in the deployment area. 
Each segment is delimited by two random points, 
and any two consecutive segments share one random point
by setting the waypoints' coordinates $(x_{W_i}, y_{W_i}, h)$ for each $W_i$, $i = 0, \ldots, n-1$
(see Fig.~\ref{fig:path_random}).
When all the waypoints are generated, the mission starts.
Once the first waypoint of \st is reached, 
the drone/rover starts to send message beacons according to its current position
by converting GPS coordinates in local Cartesian $(x,y)$ positions.
This process continues until the \ma reaches the last waypoint of \st.
When the mission is accomplished, the \ma comes back to \home.

\revision{
Regarding the \gd, in our experiments, we set its antenna on the tripod placed at \home, lying on two different planes, as sketched in Fig.~\ref{fig:setting} (left and right).
According to the \dw's Datasheet document~\cite{www-decawave2}, there are three planes in the spherical space with respect to the antenna's center, i.e., $xz$, $xy$, and $yz$.
Each plane experiences a different radiation pattern.
In Fig.~\ref{fig:setting}, we show the \gd's antenna when it is in the $xz$ plane (bottom left) and in the $yz$ plane (bottom right).
We denote as \emph{vertically placed} an antenna that lies in the $xz$ plane, and as \emph{horizontally placed} an antenna that lies in the $yz$ plane.
The drone's antenna is always vertically placed in the $xz$ plane (Fig.~\ref{fig:setting}, top left and top right) keeping the UWB transceiver at the bottom for guaranteeing the most available free space.   
For simplicity, we indicate the first configuration (Fig.~\ref{fig:setting}, left), where the two antennas lie on the same side, but in opposite direction, with vertical-vertical (VV); whereas we refer to the other configuration (Fig.~\ref{fig:setting}, right) with vertical-horizontal (VH).}

\section{Antenna Analysis}\label{sec:antenna-analysis}
In this section,
we recap the UWB technology, report the \dw's technical datasheet information,
and analyze the experimental data.

\subsection{The Ultra Wide Band Technology}
UWB is a promising radio technology that can use a very low energy level for short-range, high-bandwidth communications over a large portion of the radio spectrum.
Nowadays, its primary purpose is in the field of location discovery and device ranging. 
Differently from both Wi-Fi and Bluetooth, UWB is natively more precise and accurate, uses less power and, as production of UWB chips blows up over time, holds the promise of a lower price point.
Moreover, UWB offers relative immunity to multipath fading.

In this paper we rely on a kit of DWM1001 UWB antennas produced by \dw.
According to \dw's datasheet document~\cite{www-decawave}, those antennas provide $10 \unit{cm}$ accuracy for the measurements.
Moreover, those chips have a $6.5 \unit{GHz}$ center frequency, and have a point-to-point range up to $60 \unit{m}$ in a line-of-sight (LoS) scenario and up to $35 \unit{m}$ in a non line-of-sight (NLoS) scenario.
Although the DWM1001 chip transmitting power is set to $-41.3 \unit{dBm/MHz}$, and the typical receiver sensitivity is $-93 \unit{dBm/500 MHz}$~\cite{www-decawave}, the received power is influenced by the antenna polarization.

\subsection{Datasheet Antenna Information}
\begin{figure}[htbp]
	\centering
	\subfloat[$xz$-plane.]{%
		\centering
		\def\svgscale{1.425}
		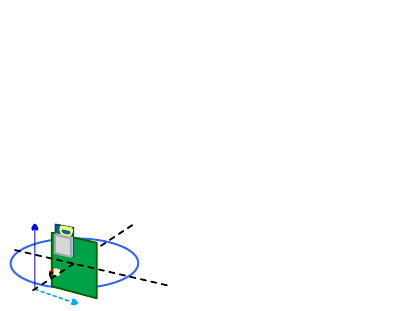
		\label{fig:dwm1001-pattern-v}
	}
	\subfloat[$yz$-plane.]{%
		\centering
		\def\svgscale{1.425}
		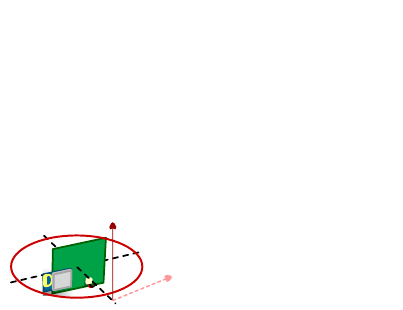
		\label{fig:dwm1001-pattern-h}
	}
	\caption{The DWM1001 radio pattern: $\unit{dBm}$ vs angle.}
	\label{fig:dwm1001-pattern}
\end{figure}

Fig.~\ref{fig:dwm1001-pattern} shows the antenna radiation patterns
of the UWB antennas according to \dw's \revision{document (\emph{Datasheet for the DWM1001C}, Tab.~12~\cite{www-decawave2})}
for different configurations.
The solid dark line of Fig.~\ref{fig:dwm1001-pattern-v} 
shows that it is possible to  obtain  the same gain in all the directions  in the $xz$-plane
when an antenna vertically placed  (i.e., on $xz$-plane) is observed by another antenna which 
shares the same {\em vertical} orientation ({\em $\Phi$ polarization}), i.e., they are concordant.
We recreate this situation by implementing the antennas as VV. 
Thus, we expect that the VV configuration   experiences the same gain at different angles, at least
when the two antennas are at the same height. 

Both the dashed lines of Fig.~\ref{fig:dwm1001-pattern-v}
and Fig.~\ref{fig:dwm1001-pattern-h}
refer to the VH configuration  because
they
show the gain 
when an antenna is observed by another antenna  {\em perpendicularly} oriented ({\em $\Theta$ polarization}), 
i.e., they are discordant.
Both the dashed lines
have nulls at certain angles that can limit the gain and
can introduce ``holes'' and  ``bubbles'' in the pattern.
Thus, we expect that the VH configuration experiences different gains at different angles, and we also expect a relevant variability given that the dashed lines of
Fig.~\ref{fig:dwm1001-pattern-h} and Fig.~\ref{fig:dwm1001-pattern-v} are different,
although the relative position of the antennas seems to be the same.
Analyzing these technical data,  
it seems that the gain is omnidirectional at least  when the two antennas are
placed as VV.

\begin{table}[htbp]
	\caption{The DW1000 elevation gains (in $\unit{dBm}$) at $6.2 \unit{GHz}$.}
	\label{tab:dw1000-elevation}
	\centering
	\begin{tabular}{c|c|cc}
		& & $\Theta$ (discordant) & $\Phi$ (concordant) \\
		\hline
		\multirow{2}{*}{$yz$-plane} & peak & $0.30$ & $2.92$ \\
		& average & $-6.99$ & $-3.04$ \\
		\multirow{2}{*}{$yz$-plane} & peak & $0.26$ & $1.39$ \\
		& average & $-5.74$ & $-3.90$
	\end{tabular}
\end{table}

We have not found, for DWM1001, 
any data which correlates the gain and the polar angle.
However, in a document of a former antenna model called DW1000~\cite{www-decawave}, 
\dw gives the gain values (reported in Tab.~\ref{tab:dw1000-elevation}) for an antenna vertically placed (as 
in Fig.~\ref{fig:dwm1001-pattern-v}) in an anechoic chamber.
We report these values just to confirm the not negligible impact of the elevation\footnote{Note also that the 3D antenna pattern is completely defined if we know its behavior in 3 planes: $xz$, $xy$, and $yz$.}:
the 3D radiation pattern is far from being a sphere, with the same gain in all the directions. 

\begin{figure}[htbp]
	\centering
	\subfloat[$xz$-plane.]{%
		\centering
		\def\svgscale{0.45}
		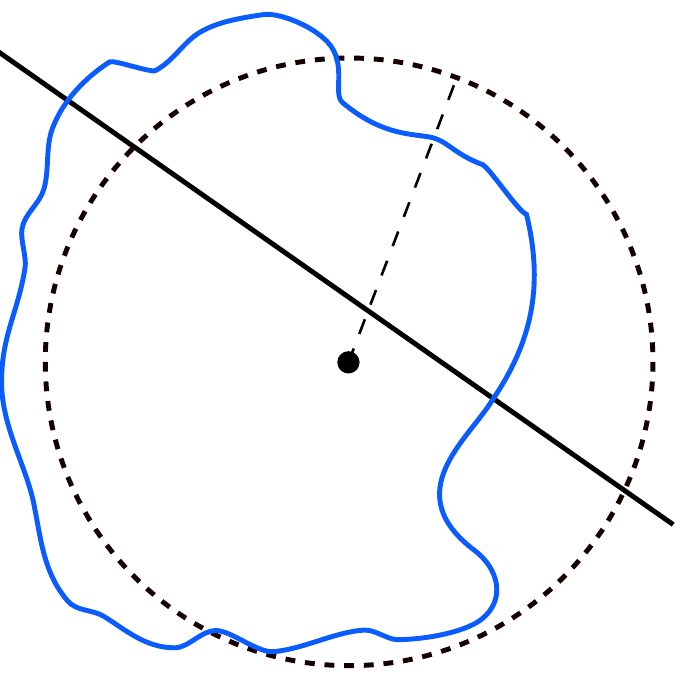
		\label{fig:cases_azimuth}
	}
	\subfloat[$yz$-plane.]{%
		\centering
		\def\svgscale{0.45}
		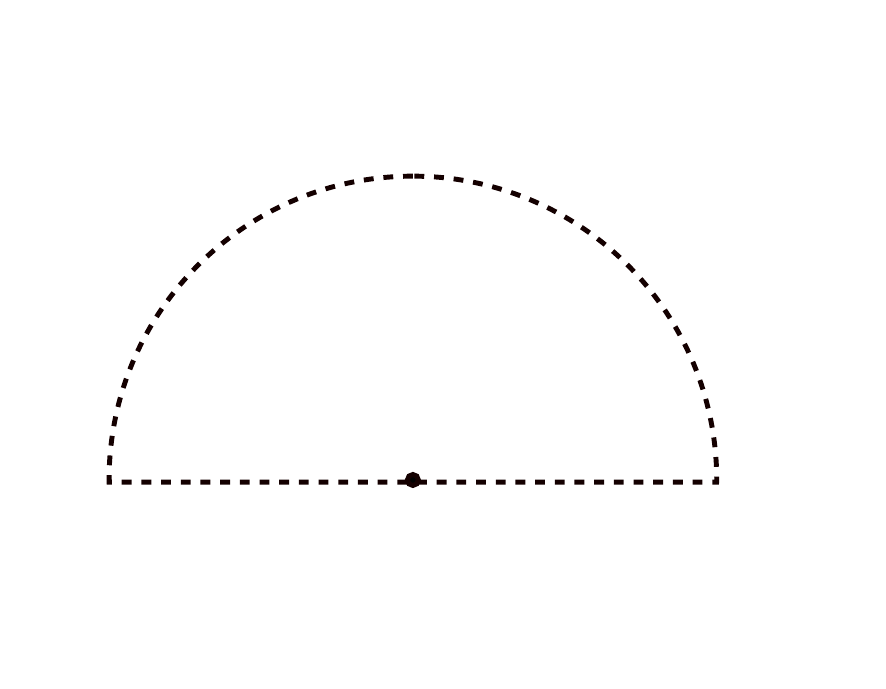
		\label{fig:cases_polar}
	}
	\caption{The ideal (dashed) and actual (solid) antenna radiation profile in $xz$-plane and $yz$-plane.}
	\label{fig:cases}
\end{figure}
Then, we conjecture that, when the \ma is a drone, the 3D antenna pattern is highly irregular 
and it can be sketched as a nibbled apple (Fig.~\ref{fig:cases}).
Therefore, when such an antenna shape is projected on the ground,
holes and  bubbles can be found.
We speculate that the altitude is the main cause of the  pattern irregularity.
So, we conjecture that when the two antennas are both on the ground (rover as \ma)
in VV configuration, the gain is the same in almost all the directions.

\subsection{Experiments for Antenna Radius}\label{sec:antenna_parameters}
In our experiments, we wish to characterize the 2D antenna pattern observing
the range of values of its radius.
The \ma starts at \home in $(0, 0)$ 
at different altitudes $h=\{0, 10, 20\} \unit{m}$.
As explained in Sec.~\ref{sec:exp-setup}, the \ma traverses the deployment area with $n$ segments that aim
to cross the receiving shape of the \gd.
Moving along each random segment, the \ma continuously broadcasts its current $(x, y)$ position, 
and the \gd registers the first and the last heard \es sent by the \ma.
Since we know in advance the \home position of the \gd, 
we can compute the actual 2D radius for each detected \e.
Note that we observe the radius on the ground.
It is important to recall that the beacons are sent at regular intervals of time,
and so the observed radii have an intrinsic error of at most \iw.
In this paper, we fix \iw to $0.40\unit{m}$ since
we have experimentally observed this  value, which clearly depends on the speed of the \ma.
From the collected \es  during the same experiment, 
we compute the mean $\mu$ and the standard deviation $\sigma$
of the set of observed radii. Then,  we apply the goodness-of-fit method in order to assess
whether a given distribution is suitable to the built data-set\footnote{
Fixed a  distribution and a set of categories, 
we determine if there is a significant difference between the expected and observed 
frequencies in one or more categories by using the chi-squared test.}. 
We repeat the experiments for the two antenna configurations (VV and VH) and for different altitudes.

We start reporting in Tab.~\ref{tab:exp-distri-h0-v} 
the results of the first experiment with the rover (i.e., $h=0\unit{m}$)
and VV configuration.
According to \dw's datasheet (solid line in Fig.~\ref{fig:dwm1001-pattern-v}), 
we expect an almost uniform radius in the experiments, 
at least when the two antennas are at the same height.
%
We observed $38$ different \es,  with mean $\mu=97.10\unit{m}$ and $\sigma=39.74\unit{m}$, and also
with $\min=32.14\unit{m}$ and $\max=162.40\unit{m}$.
According to the Pearson $\chi^2$-test, at $h=0\unit{m}$, the radii of the VV configuration have a uniform distribution.
As we will see, this is the only configuration with uniform distribution of the radii. 
Therefore, we agree with \dw that this configuration is somehow special. 
However, the radius cannot be considered really constant.

\begin{table}[htbp]
	\caption{$h=0\unit{m}$, VV}
	\label{tab:exp-distri-h0-v}
	\centering
	\begin{tabular}{c|cc|ccc}
		class & \multicolumn{2}{|c} {radii (in m)} & \multicolumn{3}{|c}{frequencies} \\	
		\# & from & to & observed & uniform ($U$) & normal ($N$) \\
		\hline
		1 & 32.14  &  58.34  & 8  & 0.08 & 3.14 \\
		2 & 58.34  &  84.54  & 7  & 0.01 & 0.13 \\
		3 & 84.54  &  110.74 & 8  & 0.08 & 0.34 \\
		4 & 110.74 &  136.94 & 5  & 0.69 & 1.06 \\
		5 & 136.94 &  163.14 & 10 & 1.06 & 8.15 \\
		\hline
		& \multicolumn{3}{c} {likelihood}  & 0.75 & 0.01
	\end{tabular}
\end{table}

Then, in Tab.~\ref{tab:exp-distri-h10-h} we report a second experiment at altitude $h=10\unit{m}$, 
with VH configuration.
We observed $28$  different \es out of $38$, thus confirming several null angles. 
The observed radii have mean $\mu=66.34\unit{m}$, $\sigma=22.81\unit{m}$, 
$\min=16.52\unit{m}$, and $\max=121.66\unit{m}$.
According to the Pearson $\chi^2$-test, at $h=10\unit{m}$, the radii of the VH configuration have a normal distribution.

\begin{table}[htbp]
	\caption{$h=10\unit{m}$, VH}
	\label{tab:exp-distri-h10-h}
	\centering
	\begin{tabular}{c|cc|ccc}
		class & \multicolumn{2}{|c} {radii (in m)} & \multicolumn{3}{|c}{frequencies} \\	
		\# & from & to & observed & uniform ($U$) & normal ($N$) \\
		\hline
		1 & 16.52 & 41.52  & 5  & 1.68  & 0.68 \\
		2 & 41.52 & 66.52  & 9  & 0.00  & 0.14 \\
		3 & 66.52 & 91.52  & 11 & 0.52  & 0.07 \\
		4 & 91.52 & 141.52 & 3  & 12.23 & 0.15 \\
		\hline
		& \multicolumn{3}{c} {likelihood}  & 0.01 & 0.90
	\end{tabular}
\end{table}

Finally, in Tab.~\ref{tab:exp-distributions} we summarize the statistic distributions 
that fit  the observed experimental radii.
For each distribution, i.e., Uniform ($U$) and Normal ($N$), 
we give the observed $\mu$ and $\sigma$, and the likelihood.
Except for the rover in VV, all the experiments show
that the radius most likely follows a normal distribution, but with a large $\sigma$. 
It is worthy to note that increasing $h$, the mean of the radii decreases.
The mean decreases faster with VV
while with VH it remains quite stable (see Tab.~\ref{tab:exp-distributions}).
The values of the radii are generally more concentrate with VH than VV. 

\begin{table}[htbp]
	\caption{The radii (in m) distribution with its parameters $D(\mu,\sigma)$ and its likelihood $p$.}
	\label{tab:exp-distributions}
	\centering
	\begin{tabular}{c|cc}
		& VV & VH \\
		\hline
		$h=0\unit{m}$ & $U(97.10, 39.74); 0.75$  & $N(63.58, 33.01); 0.40$  \\
		$h=10\unit{m}$ & $N(84.97, 31.70); 0.81$  & $N(66.34, 22.81); 0.90$  \\
		$h=20\unit{m}$ & $N(62.91, 34.06); 0.83$ & $N(57.69, 24.91); 0.82$  \\
	\end{tabular}
\end{table}

We conclude that, oppositely to our conjecture, 
VH seems better than VV and the radii obtained with a drone 
seem more concentrated than those obtained with a rover. 
Marginally, let us point out that organizing a localization mission  is easier with a drone than with a rover 
because the drone is faster and less attention has to be paid to the terrain.
Although the results are different from what we expected,
we continue our investigation in localization algorithms accuracy. 
Thus, we use the results reported in Tab.~\ref{tab:exp-distributions} to generate a large  synthetic set of \es that 
fit the estimated parameters of the radii distributions for testing the different algorithms surveyed in Sec.~\ref{sec:algorithms}. 

From now on, we refer to the radius reported in Tab.~\ref{tab:exp-distributions}
as the {\em observed radius} $r=\mu$, 
while to the \dw declared radius as the {\em manufacturer radius} $r=r_0$.

\section{Range-free Comparative Evaluation}\label{sec:simulation}
In this section, we compare all the RF localization algorithms using first the 
set of synthetic \es, and then the set of real \es collected during the experiments.

Our goal is to analyze the localization error and 
the percentage of unsuccessful localizations of \drf, \xiao, \lee, and \drfe.
From now on, with height $h=\{0,10,20\}\unit{m}$ and antenna configurations \{\,VV, VH\,\} we refer to a particular scenario.
For each simulated scenario,
we run $200$ localizations generating at random
three \es, $B_1$, $B_2$, and $B_3$,
with the distribution and the parameters of the simulated scenario derived in Sec.~\ref{sec:antenna_parameters}
from the observed \es (see Tab.~\ref{tab:exp-distributions}).
For each triple,
we invoke the four algorithms using either
the {observed radius} $r=\mu$ used to generate the \es  or 
the {manufacturer radius} $r=r_0$.
In the former case, we test the performance of
algorithms when they receive in input the actual radius,
but still, the \es can be affected by the antenna irregularity (i.e., $\sigma$).
In the latter case, we test the performance of
algorithms when they receive in input a completely different radius  from the actual one.

Reinterpreting the constraints to improve the accuracy given in~\cite{bettisorbelli2018accuracy},
the three selected \es, i.e., $B_1$, $B_2$, and $B_3$
we use for localizing the \gd should satisfy two constraints: the minimum distance $\rmin = 60 \unit{m}$ and the minimum angle $\amin=20 \unit{deg}$ between them.
The constraint $\rmin$ means that the distances $d(B_1,B_2)$, $d(B_2,B_3)$, and $d(B_3,B_1)$, must be at least as long as $\rmin$.
The $\amin$ constraint means that the three angles $\alpha_1 = \angle{B_3B_1B_2}$, $\alpha_2 = \angle{B_1B_2B_3}$, and $\alpha_3 = \angle{B_2B_3B_1}$, must be at least as large as $\amin$.
From a geometrical point of view, these two constraints guarantee that the three selected \es are sufficiently apart each other, thus avoiding the construction of degenerated triangles. 
Therefore, in the experiments, we discard any triple of  \es that does not satisfy $\rmin = 60 \unit{m}$ and $\amin=20 \unit{deg}$.
We repeat the \e extraction until we find three suitable \es.

We compare the RF algorithms under two metrics; the \textit{localization error}, defined as the Euclidean distance between the actual \gd's position and the estimated one outputted by the algorithms, and the \textit{percentage of unlocalized}.
Concerning the first metric, 
we report the localization error resumed into a 
boxplot that highlights the median (horizontal line), the average value (solid circle),
and the data between the first $Q_1$ and the third quartile $Q_3$ (box).
Additionally, the extremes of the whiskers represent the $Q_1 - 1.5~\text{IQR}$ and $Q_3 + 1.5~\text{IQR}$, respectively, where the
interquartile range (IQR) is defined as $\text{IQR} = Q_3 - Q_1$.
Lastly, about the second metric,
an unsuccessful localization is an application of the algorithm which does not return any constrained area
or in general any geometrical intersection
where the \gd can reside, i.e., the \gd remains unlocalized.
This mainly happens for the RF radius-based algorithms  when the radius is under-estimated.

\subsection{Algorithms Comparison Results}
In this section, we evaluate the performance of the RF algorithms.
We start considering a synthetic set of \es  with an average radius equal to the observed radius $\mu$, but very small standard deviation, emulating an ideal model.
This experiment is to support the observation that a high accuracy is possible when the antenna is almost isotropic.
Moreover, we discuss the performance of the RF algorithms 
on the synthetic set of \es generated according to the observed distributions 
in Tab.~\ref{tab:exp-distributions}. These experiments, as those that use the manufacturer radius, show poor accuracy.
In all the experiments, the impact of using a rover 
or a drone
is considered.
Finally, we report the comparison between the performance of \drf and \drfe.

\begin{figure}[htbp]
	\centering
	\subfloat[Error: $r=\mu$.]{%
		\centering
		\includegraphics[scale=0.65]{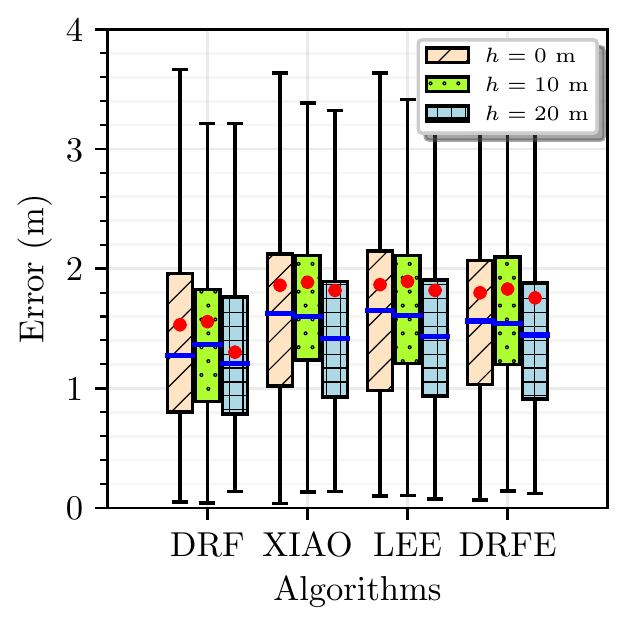}
		\label{fig:error_ideal_model_mu}
	}
	\subfloat[Error: $r=r_0$.]{%
		\centering
		\includegraphics[scale=0.65]{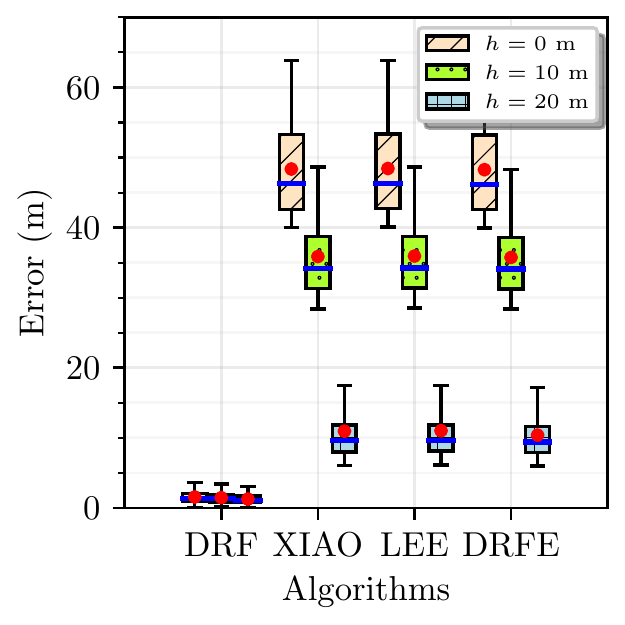}
		\label{fig:error_ideal_model_r0}
	}
	\subfloat[Unlocalized: $r=\mu$.]{%
		\centering
		\includegraphics[scale=0.65]{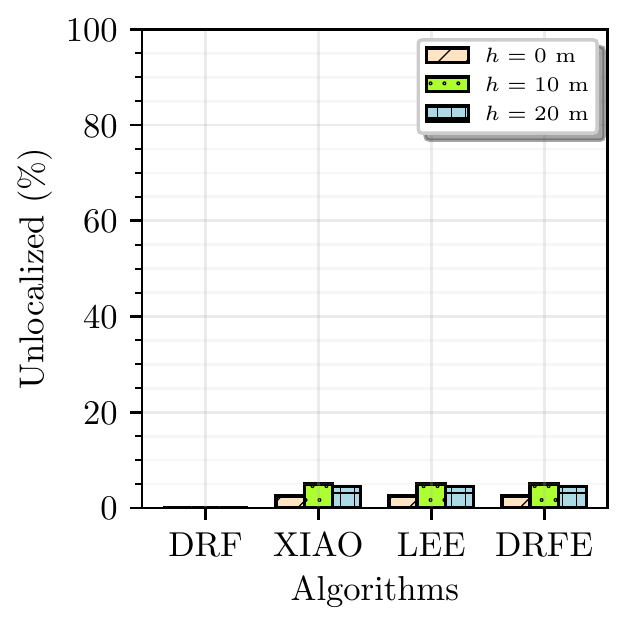}
		\label{fig:unlocalized_ideal_model_mu}
	}
	\subfloat[Unlocalized: $r=r_0$.]{%
		\centering
		\includegraphics[scale=0.65]{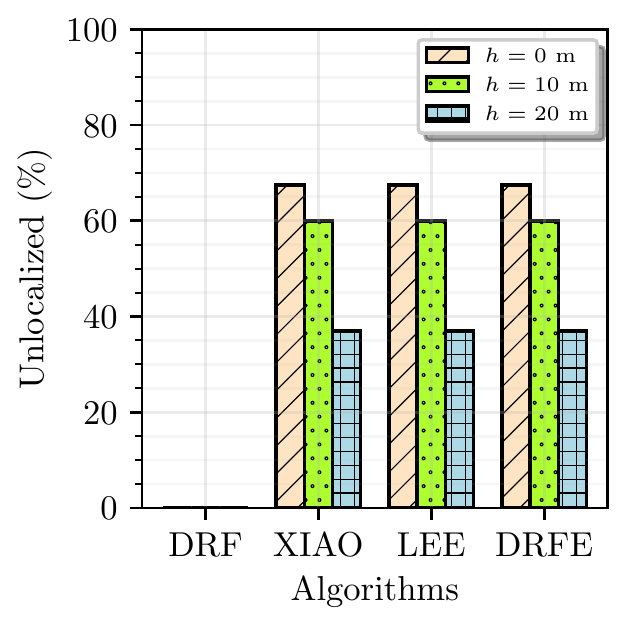}
		\label{fig:unlocalized_ideal_model_r0}
	}
	\caption{The impact of the radius on the ideal model.}
	\label{fig:plot_error_ideal_model}
\end{figure}

\paragraph{Under the Ideal Model}
Let us start considering the nearly ideal model in Fig.~\ref{fig:plot_error_ideal_model} in which  the \es 
are generated with the observed radius $\mu$  given in Tab.~\ref{tab:exp-distributions} 
for the VV configuration,  
but selecting $\sigma$ equal to $1$.
We evaluate the algorithms simulating an almost omnidirectional antenna. 
Since the dispersion is low, the radius can be considered almost constant.
As expected, all the radius-based algorithms perform well when they use  the observed radius $r=\mu$, 
which  is the same radius used to generate the \es.
In such a case, the localization error is on the order of a couple of meters or less
for all the algorithms\footnote{Please note that the scale of $y$ axis in Fig.~\ref{fig:error_ideal_model_mu}
is zoomed with respect to the scale of $y$ axis in Fig.~\ref{fig:error_ideal_model_r0}}. \drf is slightly better than the other algorithms at any altitude,
while \drfe is worse than \drf.
The number of unlocalized \gds is very small.
However, the performance of the radius-based algorithms drastically drops down 
when the algorithms run using the manufacturer radius $r_0=60 \unit{m}$, 
while the \es have been generated with the observed radius.
The percentage of unsuccessful localizations is extremely large.

These results confirm what we said in Sec.~\ref{sec:algorithms}:
to obtain an accurate localization, not only the  antenna must be of good quality (i.e., with small $\sigma$), but also the observed radius $\mu$ must be exactly known by the algorithm. 
The results show that the error due to the use of the radius $r_0$ decreases when $h$ increases because
decreases the difference between the observed $\mu$ and the manufacturer $r_0$ radii.
Alongside, note that  since \drf is radius-free, the performance of  \drf is not influenced
by the radius selection, 
as shown in Fig.~\ref{fig:error_ideal_model_r0}.
\drf only requires an omnidirectional  antenna. 

\begin{figure}[htbp]
	\centering
	\subfloat[Error: VV, $r=\mu$.]{%
		\centering
		\includegraphics[scale=0.65]{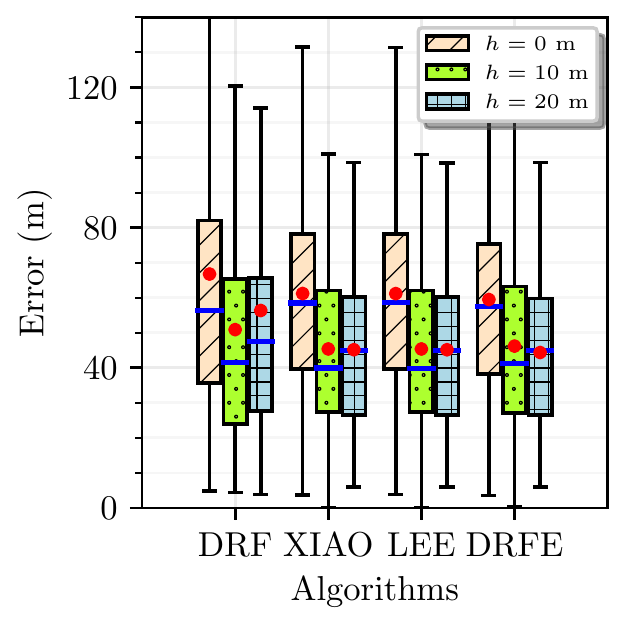}
		\label{fig:error_VV_mu}
	}
	\subfloat[Error: VH, $r=\mu$.]{%
		\centering
		\includegraphics[scale=0.65]{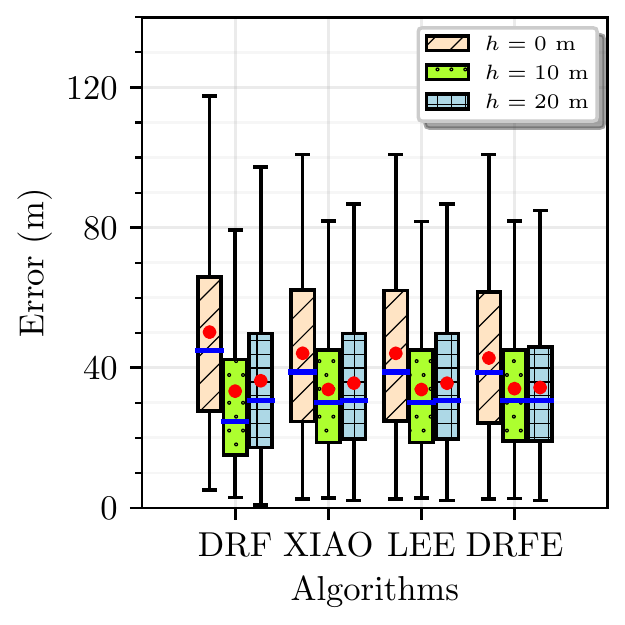}
		\label{fig:error_VH_mu}
	}
	\subfloat[Error: VV, $r=r_0$.]{%
		\centering
		\includegraphics[scale=0.65]{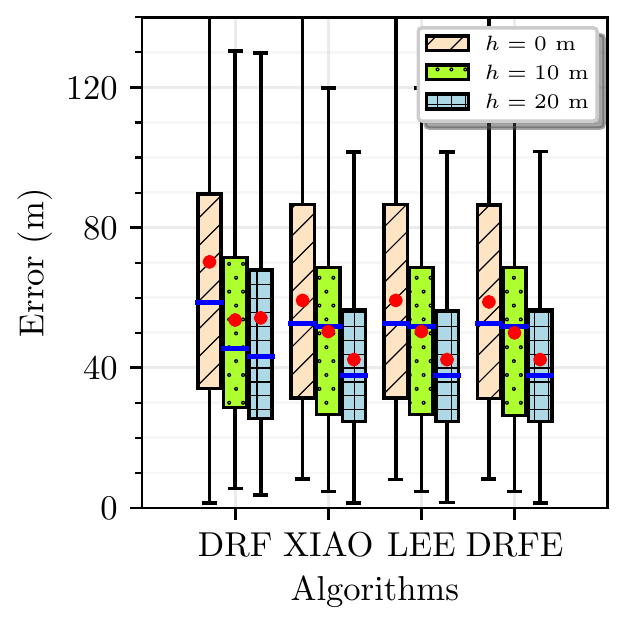}
		\label{fig:error_VV_r0}
	}
	\subfloat[Error: VH, $r=r_0$.]{%
		\centering
		\includegraphics[scale=0.65]{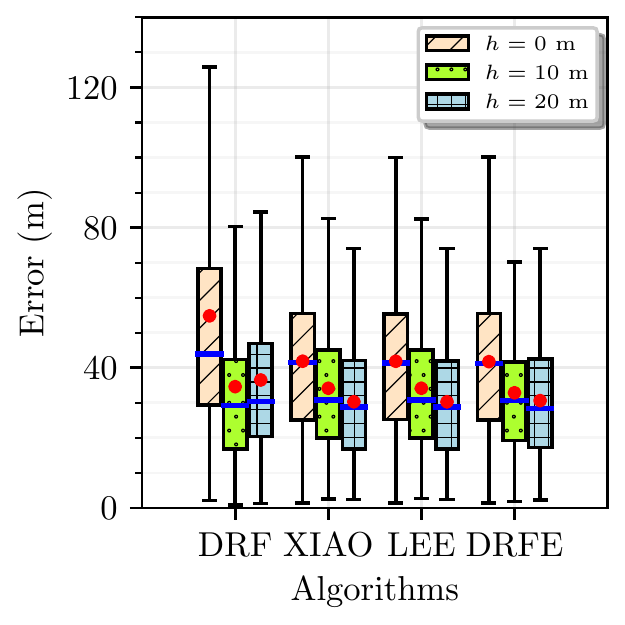}
		\label{fig:error_VH_r0}
	}
	
	\subfloat[Unlocalized: VV, $r=\mu$.]{%
		\centering
		\includegraphics[scale=0.65]{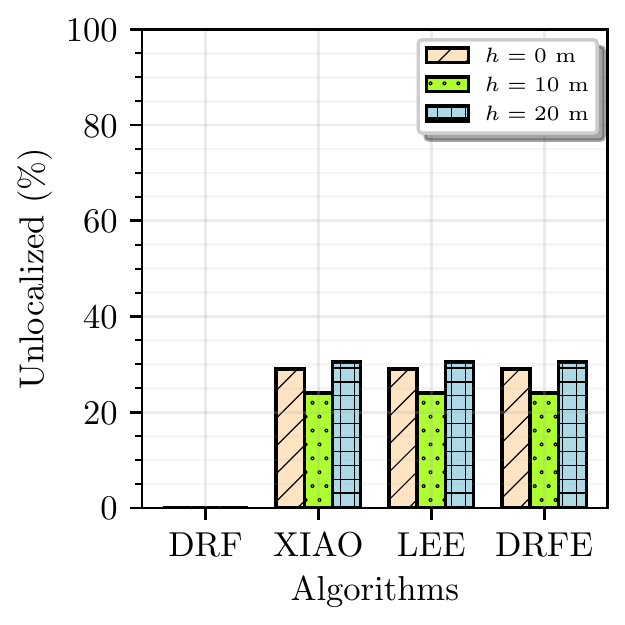}
		\label{fig:unlocalized_VV_mu}
	}
	\subfloat[Unlocalized: VH, $r=\mu$.]{%
		\centering
		\includegraphics[scale=0.65]{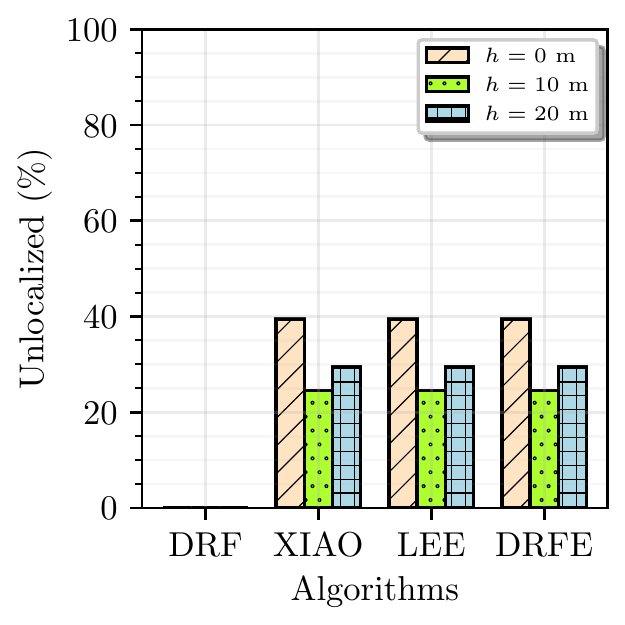}
		\label{fig:unlocalized_VH_mu}
	}
	\subfloat[Unlocalized: VV, $r=r_0$.]{%
		\centering
		\includegraphics[scale=0.65]{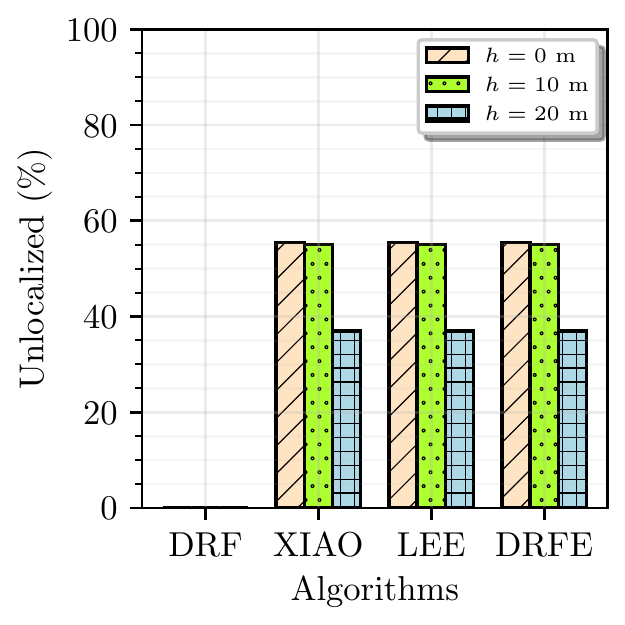}
		\label{fig:unlocalized_VV_r0}
	}
	\subfloat[Unlocalized: VH, $r=r_0$.]{%
		\centering
		\includegraphics[scale=0.65]{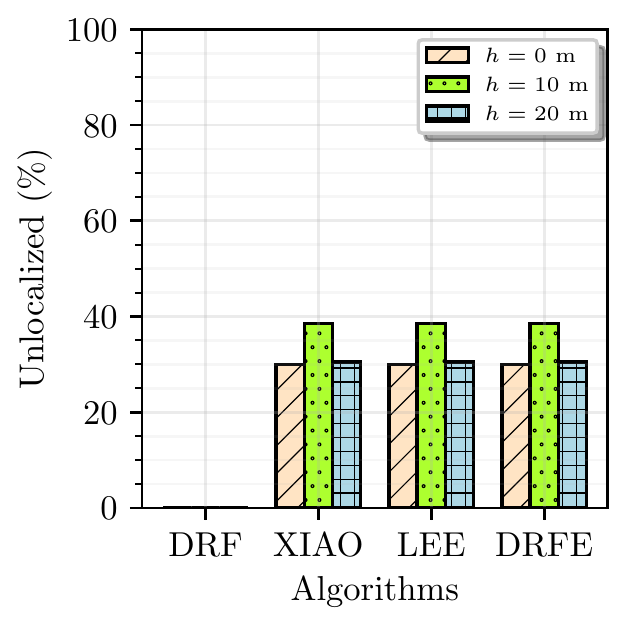}
		\label{fig:unlocalized_VH_r0}
	}
	\caption{Comparisons between all the algorithms in the studied scenarios.}
	\label{fig:plot_error_unlocalized}
\end{figure}

\paragraph{Using the Synthetic Endpoints Set}
Figs.~\ref{fig:error_VV_mu},~\ref{fig:error_VH_mu},~\ref{fig:unlocalized_VV_mu}, and~\ref{fig:unlocalized_VH_mu} 
show the results
when both the \es and the algorithms use the observed radius 
in Tab.~\ref{tab:exp-distributions}.
Figs.~\ref{fig:error_VV_r0},~\ref{fig:error_VH_r0},~\ref{fig:unlocalized_VV_r0}, and~\ref{fig:unlocalized_VH_r0} 
instead show the results when the \es are generated using the observed radius and the algorithms run with the manufacturer radius. This is what happens in practice whenever the end-user  implements the RF radius-based algorithms using the \dw datasheet radius on our test-bed.
The errors of the algorithms are
compared at different altitudes, antenna configurations, and radii.
For all the algorithms, the average error
is  large. 
The worst error occurs with the VV configuration of the antennas
and $h=0\unit{m}$:
in such a scenario, the \es follow a uniform distribution.
In general, the VH makes an error smaller than the VV   
probably because the radius that generated the \es for VH is less dispersed (i.e., $\sigma$ is smaller) than that for VV.
The three radius-based algorithms, \xiao, \lee, and \drfe, exhibit the 
same performance. They  are inspired by slightly different ideas, but
they actually act  the same. 

The \drf algorithm experiences  the worst average error,
but the error is only slightly more than that of the radius-based algorithms.
The  whisker of the largest error of \drf is the longest whisker
among all the algorithms
probably because \drf finds a localization also in extreme cases
when other algorithms return an unsuccessful localization.
The error in \drf decreases when the altitude increases,
moreover the error of the VV configuration is worse than that of the VH one. 

The localization error is almost the same
regardless of the adopted radius (observed or manufacturer). 
As witnessed by comparing
the whiskers of the boxplots in Figs.~\ref{fig:error_VV_mu} and~\ref{fig:error_VV_r0}
(resp., by Figs.~\ref{fig:error_VH_mu} and~\ref{fig:error_VH_r0}),
the localization error of the radius-based algorithms,
when they use the manufacturer radius $r_0$,
is slightly worse than when they use the observed radius $\mu$.
Instead, the knowledge of $\mu$ reduces percentage of unlocalized \gds
as illustrated in Figs.~\ref{fig:unlocalized_VV_mu} and~\ref{fig:unlocalized_VV_r0}
for VV.
The improvement in Figs.~\ref{fig:unlocalized_VH_mu} and~\ref{fig:unlocalized_VH_r0} is  weaker for VH
than for VV because for VH the difference between the observed radii $\mu=\{63.58,66.34,57.60\}\unit{m}$ and $r_0=60\unit{m}$ is  smaller than VV.

In conclusion, the average localization error for VV and VH is $50 \unit{m}$ and $30 \unit{m}$, respectively,
regardless of if the algorithm knows the true radius used to generate the \es or not.
The knowledge of the radius used to generate the \es by the algorithm only matters when the
radius dispersion is small.
The take-away lesson here is that to apply a RF radius-based algorithm the antenna must be omnidirectional and its radius must be known by the algorithm.

%

\begin{figure}[htbp]
	\centering
	\subfloat[Error: VV, \drf vs \drfe.]{%
		\centering
		\includegraphics[scale=0.65]{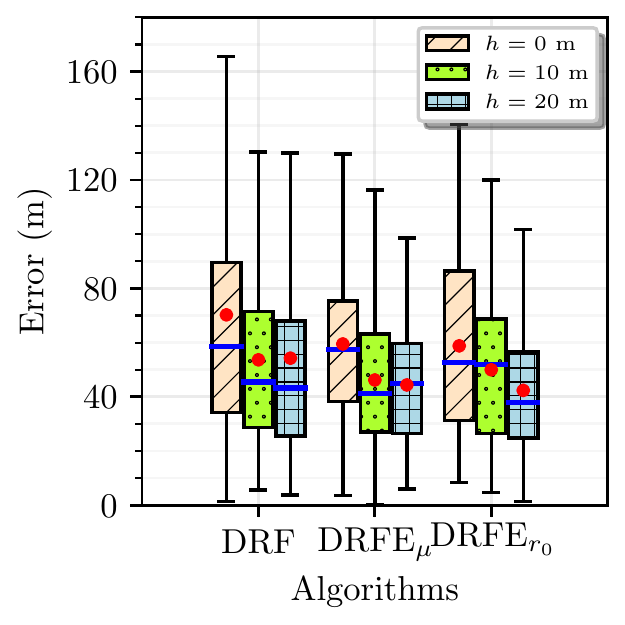}
		\label{fig:error_drf-vs-drfevo_VV}
	}
	\subfloat[Error: VH, \drf vs \drfe.]{%
		\centering
		\includegraphics[scale=0.65]{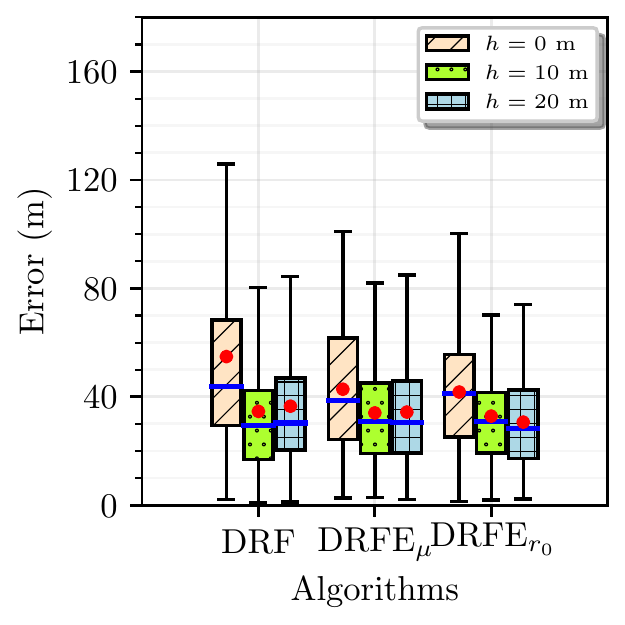}
		\label{fig:error_drf-vs-drfevo_VH}
	}
	\caption{\drf vs \drfe.}
	\label{fig:plot_error_drf-vs-drfevo}
\end{figure}

\paragraph{Between \drf and \drfe}
In Fig.~\ref{fig:plot_error_drf-vs-drfevo},
we compare the performance of \drf, 
$\drfe_{\mu}$ (\drfe with radius $\mu$), 
and   $\drfe_{r_0}$ (\drfe with radius $r_0$).
In general,
$\drfe_{\mu}$ performs better  than $\drfe_{r_0}$,
but $\drfe_{\mu}$  has always an error much greater than the error of the ideal
model (see Fig.~\ref{fig:plot_error_ideal_model}).

\begin{figure}[htbp]
	\centering
	\subfloat[Error: VV, $\mu$.]{%
		\centering
		\includegraphics[scale=0.65]{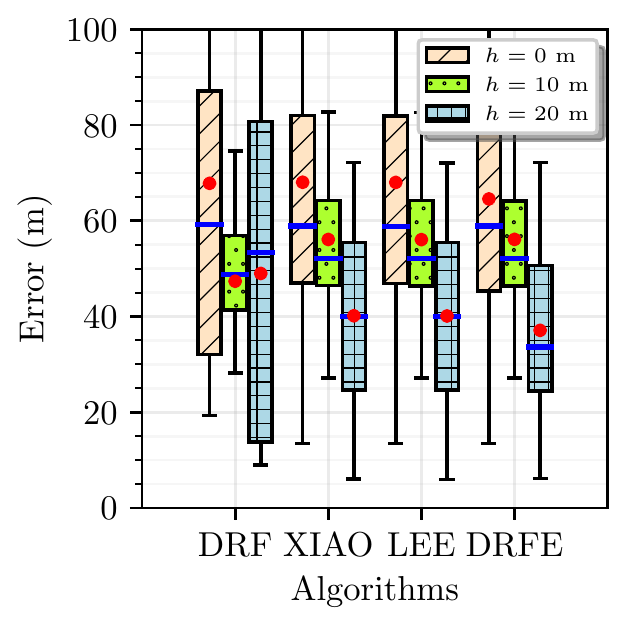}
		\label{fig:error_real_VV_mu}
	}
	\subfloat[Error: VH, $\mu$.]{%
		\centering
		\includegraphics[scale=0.65]{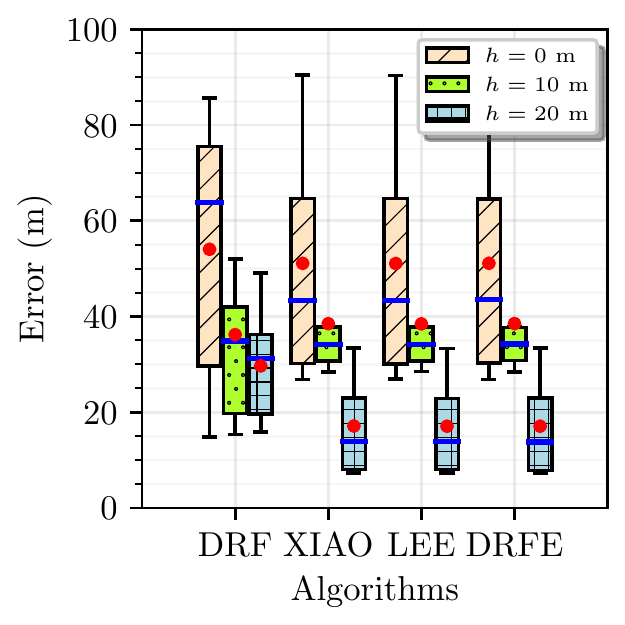}
		\label{fig:error_real_VH_mu}
	}
	\subfloat[Error: VV, $r_0$.]{%
		\centering
		\includegraphics[scale=0.65]{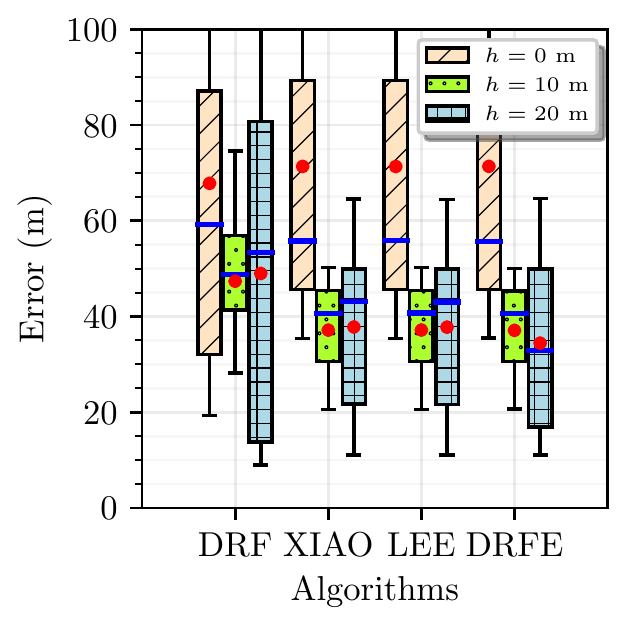}
		\label{fig:error_real_VV_r0}
	}
	\subfloat[Error: VH, $r_0$.]{%
		\centering
		\includegraphics[scale=0.65]{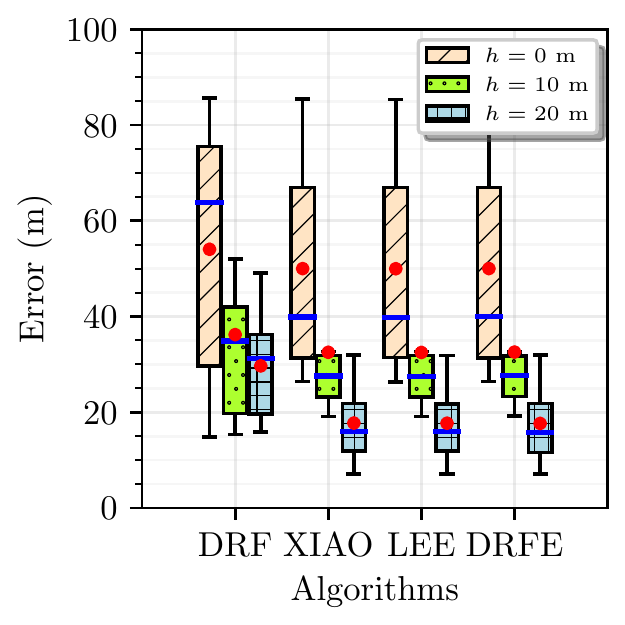}
		\label{fig:error_real_VH_r0}
	}
	\caption{Error using real data.}
	\label{fig:error_real}
\end{figure}

\paragraph{Using Real Endpoints}
In Fig.~\ref{fig:error_real}, we show the results using the real \es collected during our test-bed. 
%
For the radius-based algorithms, in VV, the results of the localization error seem to follow the trend already seen on the synthetic generated set.
In general, VV has a larger error compared to that of VH.
In VH,  the results are slightly better than that obtained on
the synthetic generated set, although the number of valid triples is small.
Indeed, the average error at $h=20\unit{m}$ is about $15 \unit{m}$ although $\sigma$ is greater than $20 \unit{m}$
when the radius $\mu$ is used. The average error increases to $20 \unit{m}$ 
when the radius $r_0$ is used. 
This can be explained with the fact that there is a much stronger correlation between the \es
that are not fully captured by the two constraints
\rmin and \amin that we imposed for the 
selection of the synthetic \es.
Finally, 
the error of \drf is worse with real \es than with  synthetic \es,
especially for the VV configuration.

\begin{figure}[htbp]
	\centering
	\def\svgscale{0.5}
	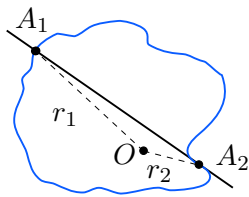
	\caption{The two radii measure by the \gd when the antenna pattern is irregular.}
	\label{fig:bubble_measurement}
\end{figure}

So far, we learned the performance of all the RF algorithms strongly relies on the radiation antenna pattern.
The simplest way to discover the radiation antenna pattern is to measure for each direction up to which distance
the \ma and the \gd are one in the range of the other.
So, in the next section, we significantly improve the localization accuracy 
exploiting the capability of the UWB antennas used in our test-bed of taking distance measurements.
Even though there are  bubbles and holes in the antenna pattern, it is always possible
to discover them if the distance from the \gd and the \ma is taken (see Fig.~\ref{fig:bubble_measurement}).

\section{Range-based Comparative Evaluation}\label{sec:range-based}
Our interest in RF algorithms lay in the fact that they are simple to implement,
do not require specialized hardware, and are scalable with respect to the number of \gds.
Also, they are immune to problems that
come from the  measurement of the 3D distance. 
Indeed, distance measurements are affected by several errors 
that depend on the adopted technology and GPS.
Technologies like WiFi or Bluetooth are much less accurate than UWB and may 
\revision{have} measurement errors of the order of tens of meters,
whereas GPS and barometer inaccuracies together with bad weather conditions can
seriously impact the drone's position.
Such combined 3D slant errors are then reflected on the ground, leading to 2D ground errors.
Although range measurements come with their troubles, given the results
of the previous section,
we decided to include distance measurements in our algorithms. 
Since in our test-bed both the \ma and \gd are equipped with UWB antennas that
are able to take distance measurements, we do not need to heavily modify our test-bed.

\begin{figure}[htbp]
	\centering
	\small
	\subfloat[RB \xiao.]{%
		\def\svgscale{0.65}
		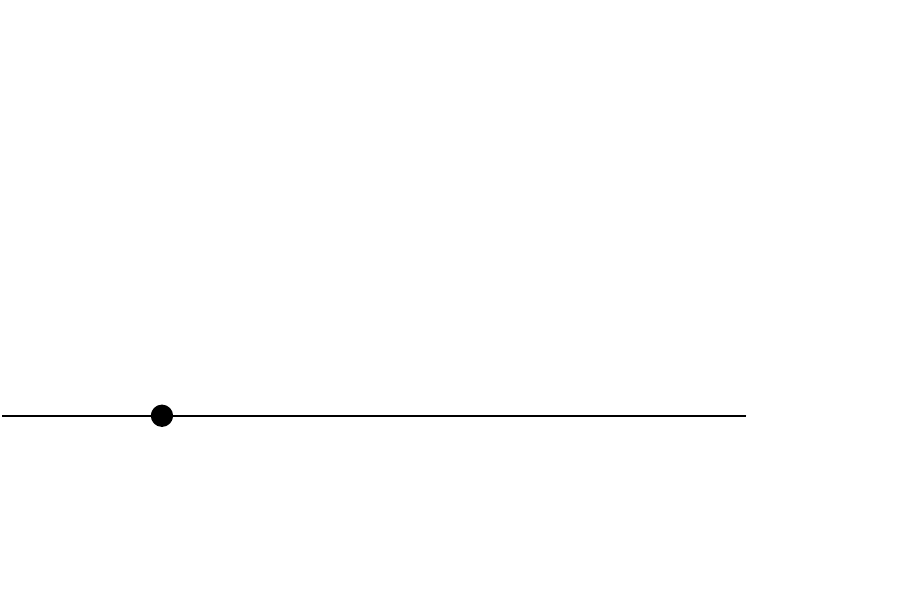
		\label{fig:model-xiao_m}
	}
	\subfloat[RB \lee.]{%
		\def\svgscale{0.65}
		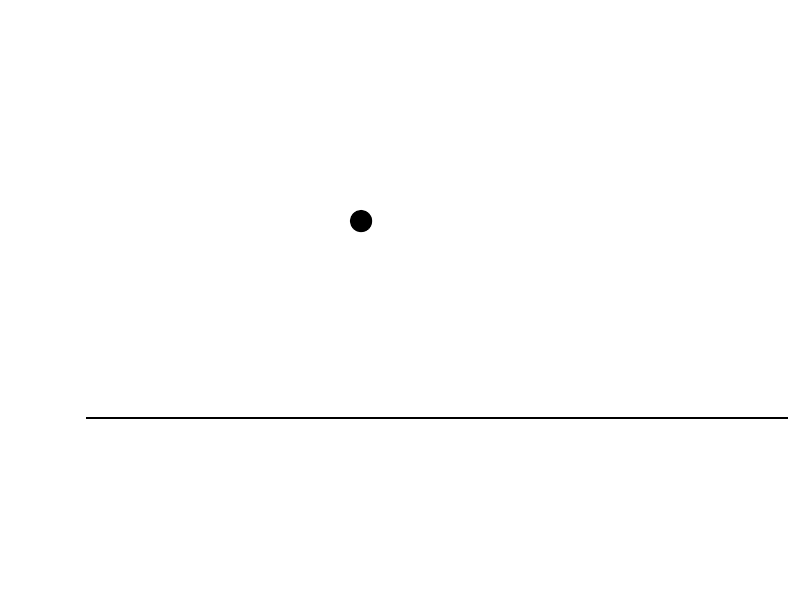
		\label{fig:model-lee_m}
	}
	\subfloat[RB \drfe.]{%
		\def\svgscale{0.65}
		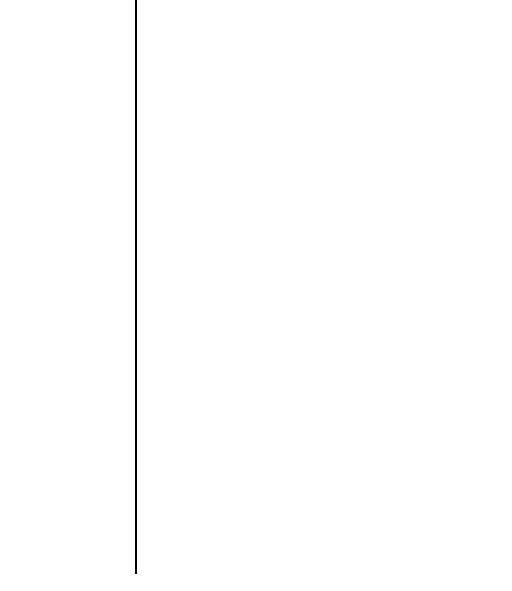
		\label{fig:model-drf_evo_m}
	}
	\caption{The \xiao, \lee, and \drfe localization algorithms using two radii.}
	\label{fig:intersection_areas_m}
\end{figure}

In the algorithm variants that we are going to propose, the actual measured distance between \ma and \gd
is used to run the algorithms
instead of using the observed or manufacturer radius.
This adaption makes the RF algorithms actually RB algorithms,
but we maintain the original algorithmic rules (e.g., intersection of annuli or circumferences).
It is worth noting that we can apply modifications only to the RF algorithms that actually
make use of the radius in their localization rules, and hence only for the radius-based ones, i.e.,
\xiao, \lee, and \drfe.
Accordingly, the simplest radius-free algorithm \drf cannot be adapted to measurements.

\paragraph{Range-based version of \xiao}
Fig.~\ref{fig:model-xiao_m} shows how the \xiao algorithm is modified.
When the \gd hears for the first time the beacon in $A_1$, the \gd measures the distance $r'_1 = d(A_1, O)$ between its position $O$ and $A_1$. 
Whenever the \ma is inside the receiving area of the \gd, the \gd still takes distance measurements neglecting the intermediate measurements.
When the last beacon $A_2$ is heard, the \gd saves the last distance $r'_2=d(A_2,O)$ between its position $O$ and $A_2$. 
As in the original version of \xiao, the \gd computes the other two points $A_0$ and $A_3$ knowing the line that interconnects $A_1$ and $A_2$ and the value of \iw. 
Then, the algorithm proceeds as before, except that it draws two circumferences of radius $r_1= r'_1+\iwd$ centered in $A_0$ and $A_1$, and two circumferences of radius $r_2=r'_2+\iwd$ centered in $A_2$ and $A_3$.
That is, the predefined observed or manufacturer radii are substituted by the distances from the \es and the \gd.
Note that the radius $r_1$ and $r_2$ are used instead of $r'_1=d(A_1,O)$ and $r'_2=d(A_2,O)$ to avoid that $O$ falls outside the intersection area in case of measurement error that underestimated the distances (see Fig.~\ref{fig:model-xiao_m}). 

\paragraph{Range-based version of \lee}
As seen for \xiao, also for \lee, the two distances $r'_1$ and $r'_2$ between the \es and the \gd are ranged.
The algorithm proceeds by drawing two annuli centered at $A_1$ and $A_2$, with outer radius equal to $r_1 = r'_1 + \iwd$ and $r_2 = r'_2 + \iwd$, and inner radius equal to $r_1 - \iw$ and $r_2 - \iw$, respectively.
The main difference with the previous version is that the radii are replaced by the distances between the \gd and the \es.
As in \xiao, we use the radii $r_1 = r'_1 + \iwd$ and $r_2 = r'_2 + \iwd$ instead of $r'_1$ or $r'_2$ to limit the risk that, due to measurement inaccuracy, $O$ falls outside the intersection area.
The width $\iw$ of the annuli is preserved.

\paragraph{Range-based version of \drfe}
We replace $r$ with $r_1=r'_1=d(A_1,O)$ and $r_2=r'_2=d(A_2,O)$.
The intersection of the two circles of radius $r_1$ and $r_2$ centered, respectively, 
at $A_1$ and $A_2$, returns $P_1$ which coincides with $O$ (assuming no measurement errors).
To repeat the construction of the RF \drfe, 
the points $P_3$ and $P_2$ are drawn in a  way similarly to $P_1$. Precisely, $P_3$ is at the intersection between a circumference
of radius $r_1$ centered in $A_0$ and a circumference of radius $r_2$ centered in $A_2$.
$P_2$ is at the intersection between a circumference
of radius $r_1$ centered in $A_1$ and a circumference of radius $r_2$ centered in $A_3$.
As a result, differently from the original version where there were three isosceles triangles
lying on the same line passing through $A_0$ and $A_3$,
here there are three  scalene triangles, i.e.,
$\triangle{(A_1A_2P_1)}$,  $\triangle{(A_1A_3P_2)}$, and $\triangle{(A_0A_2P_3)}$.
Eventually, the centroid resulting from the vertices $P_1$, $P_2$, and $P_3$
is selected as the estimation point,
as shown in Fig.~\ref{fig:model-drf_evo_m}.

\subsection{Error Analysis with Different Radii}
In this section, we argue concerning the localization error that can be obtained once the original RF versions of the algorithms have been adapted to be actually RB. 
We focus on \xiao, but a very similar analysis applies to \lee. 
Some words will be finally spent for \drfe. 

Without loss of generality, we assume that the \ma moves along a straight line along the $x$-axis (see Fig.~\ref{fig:model-xiao_m}). 
Let $O=(x_O, y_O)$ be the actual \gd's location with respect to Cartesian coordinate system with origin in $A_0=(0,0)$, and let $P=(x_P, y_P)$ be the estimated \gd's location by the algorithm. 
We fix $A_1$ and $A_2$ so as $r_1=d(A_1,O) \le d(A_2,O)=r_2$ and let $A_1=(\iw, 0)$ and $A_2=(k \iw, 0)$, where $k$ is an integer number that represents the number of times the \ma has sent the beacons from $A_1$ to $A_2$. 
Finally, let $A_3=((k+1) \iw, 0)$.

The RB \xiao relies on two different radii, i.e., $r_1$ applied to $A_0$ and $A_1$, and $r_2$ applied to $A_2$ and $A_3$, for constructing four circumferences, i.e.,
\begin{inparaenum}[(a)]
	\item $x^{2}+y^{2} > r_{1}^{2}$, \label{eq:a0} centered in $A_0$,
	\item $\left(x-\iw\right)^{2}+y^{2} \le r_{1}^{2}$, centered in $A_1$, \label{eq:a1}
	\item $\left(x-k\iw\right)^{2}+y^{2} > r_{2}^{2}$, centered in $A_2$, and \label{eq:a2}
	\item $\left(x-\left(k+1\right)\iw\right)^{2}+y^{2} \le r_{2}^{2}$, centered in $A_3$. \label{eq:a3} 
\end{inparaenum}
We also denote as $L_1$ the \emph{lune}\footnote{In plane geometry, a lune is the concave-convex region bounded by two circular arcs.} delimited by Eqs.~\eqref{eq:a0}--\eqref{eq:a1}, and as $L_2$ the lune delimited by Eqs.~\eqref{eq:a2}--\eqref{eq:a3}, also in accordance with the original \xiao algorithm.

Note that $r_1$ and $r_2$ are affected by the measurement errors~\cite{bettisorbelli2018accuracy} and for this reason we do not directly conclude that the estimated position $P$ is at the intersection of Eqs.~\eqref{eq:a1}--\eqref{eq:a2}.
Moreover, the quantity $k\iw = d(A_1, A_2)$ is influenced by the \ma's speed, and hence it cannot be considered exact as well.
Therefore, we prefer to select $P$ inside the intersection among $L_1$ and $L_2$.

The correctness of the RB \xiao algorithm is completely different from that of the original \xiao. 
Although our implementation only memorizes the first and the last measured radius, i.e., $r_1$ and $r_2$, the RB \xiao algorithm performs measuring, at each $i^{th}$ beacon $A_i=(i\iw,0)$, the current distance $r_2(i)=d(A_i,O)$.
Therefore, for each intermediate beacon $A_i$ and corresponding measurement $r_2(i)$, with $1 \le i \le k$, RB \xiao knows that $O \in (L_1 \cap L_2(i))$, i.e., where $L_2(i)$ is the lune created by the two circumferences of radius $r_2(i)$ centered in $A_i$ and $A_{i+1}$, respectively. 
Hence, although we do not implement this feature in our algorithm, after each measurement $r_2(i)$ RB \xiao could stop.
As we will see, the best moment to stop would be when $r_2(i)$ is minimum.
This is the main reason that makes RB \xiao robust to hole and bubbles: whenever it stops, the intersection area is limited.
However, a discussion about the intersection of lunes while $A_i$ varies along the $x$-axis is needed to complete the error analysis. 

\begin{figure}[htbp]
	\centering
	\hfill
	\subfloat[Tangent concordant.]{%
		\def\svgscale{0.8}
		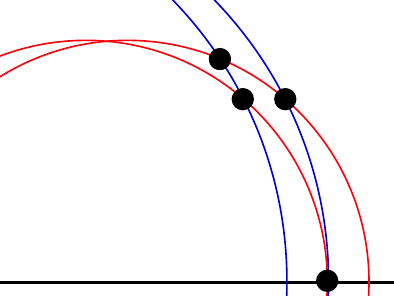
		\label{fig:curvature-concordant}
	}
	\hfill
	\subfloat[Parallel to $x$-axis.]{%
		\def\svgscale{0.8}
		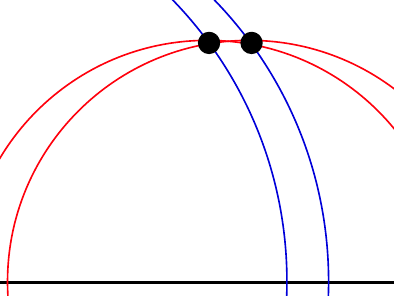
		\label{fig:curvature-0}
	}
	\hfill
	\subfloat[Tangent discordant.]{%
		\def\svgscale{0.8}
		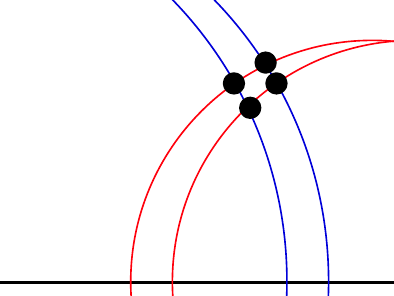
		\label{fig:curvature-discordant}
	}
	\hfill
	\subfloat[When $C$ is undefined.]{%
		\def\svgscale{0.8}
		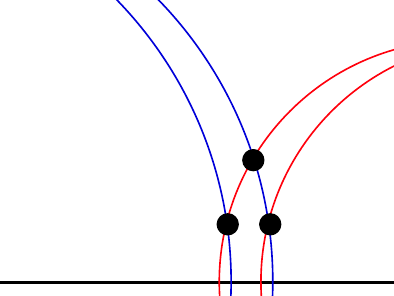
		\label{fig:xiao_k_err_3_points}
	}
	\hfill
	\caption{The possible intersection areas depending on the lunes: the solid line is the $x$-axis.}
	\label{fig:xiao_intersections}
\end{figure}

Assuming an omnidirectional antenna pattern, as \ma moves along the $x$-axis, it crosses the \gd's receiving area disk. 
The distances $d(A_i,O)$ decrease until \ma reaches the closest position to $O$ in $A_{k^*}=(k^*\iw,0)$ (precisely, $x_O \le k^*\iw < x_O+\iw$). 
Then, $d(A_{k^*},O) \approx y_O$. 
After $A_{k^*}$, the distances $d(A_i,O)$ start to increase again up to $A_3=((k+1)\iw, 0)$, where the \ma is no longer reachable from the \gd, and whose last measured distance taken from $A_2=(k\iw,0)$ is $d(A_2,O)=r_2$.
The observed distances while \ma moves in $A_i$, with $i=1, \ldots, k^*, \ldots, k$,  form  a unimodal sequence $R$ with minimum in $k^*\iw$.
As long as the radii in $R$ decrease, i.e., $x_{A_i} < x_{A_{k^*}}$, the lunes $L_1$ and $L_2(i)$ have the same \emph{curvature}\footnote{The curvature is the sign of the tangent to the curve.} (see Fig.~\ref{fig:curvature-concordant}), while when the radii in $R$ increase, i.e., $x_{A_i} > x_{A_{k^*}}$, $L_1$ and $L_2(i)$ have opposite curvature (see Fig.~\ref{fig:curvature-discordant}).
In $A_{k^*}$, the tangent to the circumferences of the lune $L_2(k^*)$ is parallel to the $x$-axis (see Fig.~\ref{fig:curvature-0}).

In a general scenario in presence of holes and bubbles, the \ma stops at $A_2$ in one of the three cases depicted in Fig.~\ref{fig:xiao_intersections}:
\begin{inparaenum}[(i)]
    \item the curvatures of $A_1$ and $A_2$ are concordant, i.e., $1 < k < k^*$ or $k^* < 1 < k$; \label{cases:1}
    \item the tangent at $A_2$ is parallel to $x$-axis, i.e., $1 < k = k^*$; \label{cases:2}
    \item the curvature $A_1$ and $A_2$ are discordant, i.e., $1 < k^* < k$. \label{cases:3}
\end{inparaenum} 
To evaluate the size of the intersection among $L_1$ and $L_2$ in these three cases, let observe that they cross each other making at most four intersections, i.e., $A$ is the intersection among Eqs.~\eqref{eq:a1}--\eqref{eq:a2}, $B$ among Eqs.~\eqref{eq:a1}--\eqref{eq:a3}, $C$ among Eqs.~\eqref{eq:a0}--\eqref{eq:a3}, and $D$ among Eqs.~\eqref{eq:a0}--\eqref{eq:a2} (see Fig.~\ref{fig:xiao_intersections}), where:

{
\begin{equation*}
	\begin{aligned}[c]
		A\!&=\!\left(\frac{r_{1}^{2}-r_{2}^{2}}{2\left(k-1\right)\iw}+\frac{k+1}{2}\iw,\sqrt{r_{1}^{2}-\left(\frac{r_{1}^{2}-r_{2}^{2}+\left(k-1\right)^{2}\iw^{2}}{2\left(k-1\right)\iw}\right)^{2}}\right) \\
		B\!&=\!\left(\frac{r_{1}^{2}-r_{2}^{2}}{2k\iw}+\frac{k+2}{2}\iw,\sqrt{r_{1}^{2}-\left(\frac{r_{1}^{2}-r_{2}^{2}+k^{2}\iw^{2}}{2k\iw}\right)^{2}}\right) \\
		C\!&=\!\left(\frac{r_{1}^{2}-r_{2}^{2}}{2\left(k+1\right)\iw}+\frac{k+1}{2}\iw,\sqrt{r_{1}^{2}-\left(\frac{r_{1}^{2}-r_{2}^{2}+\left(k+1\right)^{2}\iw^{2}}{2\left(k+1\right)\iw}\right)^{2}}\right) \\
		D\!&=\!\left(\frac{r_{1}^{2}-r_{2}^{2}}{2k\iw}+\frac{k}{2}\iw,\sqrt{r_{1}^{2}-\left(\frac{r_{1}^{2}-r_{2}^{2}+k^{2}\iw^{2}}{2k\iw}\right)^{2}}\right)
	\end{aligned}
\end{equation*}
}
Note that $A$ is stable when $A_2$ moves because it is at the intersection of the two measurements of $O$. If there were no errors, $O \equiv A$. 
It is worth noting that $y_B = y_D$ and $x_B - x_D = \iw = d(D,B)$ regardless of $k$.
Now we are in the position of clarifying the three previous cases.
Recalling that $A_1=(\iw,0)$, depending on which is the last heard beacon $A_2=(k\iw,0)$, it may occurs:

{\bf Case~\eqref{cases:1}}: The lunes have the same curvature, as illustrated in Fig.~\ref{fig:curvature-concordant}. 
The area where $P$ can be selected has width $\iw$ and height $y_C-y_A$.
In particular, if $k$ is very small, $r_2 \to r_1$, and thus $y_C \to r_1$. 
If $y_A \to 0$, the error $y_C -y_A \approx r_1$. 
As $k$ approaches $k^*$, $y_C -y_A$ quickly decreases. 
A similar error occurs when $k^*< 1 <k$. 

{\bf Case~\eqref{cases:2}}: By intersecting the coordinates $x_C=x_A$, and all the four points $A$, $B$, $C$, and $D$ are very close. The area where $P$ can be selected is very small. 

{\bf Case~\eqref{cases:3}}: Observe that $x_C$ can be re-written as $x_C \approx \left(x_{A}+\iw\right)\ \frac{k-1}{k+1}$  for $k \ge 2$.
Thus, $x_C < x_A+\iw$ for any  $k$.
Due to this and the fact that the curvatures of $L_1$ and $L_2$ are opposite, their intersection area is quite limited. 
Unfortunately, we could not find a simple formula for describing $|y_A-y_C|$ here,
but it can be computed by approximating the curves with their tangent in $A$.
Hence, the intersection of the two lunes can be inscribed in a rectangle with two sides parallel to the $x$-axis of length \iw and two vertical sides whose length depends on the angular coefficient of the tangents in $A$. 
We can only add that their lengths decreases below $2 \iw$ when $\arctan\frac{y_A}{x_A} > 1$. So, the area where $P$ can be selected only depends on $\iw$.  
When $\arctan\frac{y_A}{x_A} \le 1$, the trivial bound is $y_A$, which
cannot be very large however.  So, the size of the area where $P$ can be selected depends on $\iw$ and $y_A$. 
When $r_1 \le x_A < r_1 + \iw$, $C$ is undefined (see 
Fig.~\ref{fig:xiao_k_err_3_points}), Hence,  the vertical side of the rectangle that contain $P$  has length at most $y_A \le \sqrt{r_1^2-(r_1-\iw)^2}$. 
This leads to the same error described in \xiao when the lunes have only $3$ intersection points~\cite{xiao2008distributed}. 

In presence of irregularities, RB \xiao, as in our implementation, stops at the last heard beacon. 
The error can be large only when $A_1$ and $A_2$ fall on the same side with respect to $x_O$, and thus the associated lunes have concordant curvatures. 
In order to limit the occurrences of Case~\eqref{cases:1}, in our implementation we have forced $B_1$, $B_2$, and $B_3$, i.e., the three selected \es, to respect the \rmin and \amin constraints, putting them sufficiently apart, so as the two lunes will have opposite curvature.
Moreover, in Cases~\eqref{cases:2} and~\eqref{cases:3} the error is quite limited and only depends on \iw. 
Without irregularities, RB \xiao always falls in Cases~\eqref{cases:2} or~\eqref{cases:3}.
The same happens for the original \xiao which stops when $r_1=r_2$ and thus the curvatures are opposite. 

To conclude, the above analysis also applies for \lee. 
In \lee, the \gd belongs to the intersection of two annuli of different radii but with the same width \iw, and the error depends on the curvature of the lunes created by the intersection of annuli.
%
As regard to \drfe, $P_1$, $P_2$, and $P_3$ are the intersections of pair of circumferences, and here the lunes coincide with portions of circumferences.
Differently from \xiao and \lee, the position uncertainty depends only on the measurement errors and on \iw, and it is bounded according to~\cite{bettisorbelli2018accuracy}.

\subsection{Range-based Versions Results}
In this section, we evaluate the performance of the RB variants, and we compare
these results with those of the RF original implementation.

In reality, in our previous experiments, since the goal was to investigate the performance of RF algorithms,
we did not take any distance measurement even though our UWB
antennas were able to range measurements with good accuracy.
Indeed, we only knew in advance the \gd's \home position $(0,0)$ 
and the \ma's position with respect to \home. 
Instead, the new RB variants rely on the 3D slant distances between the \ma and the \gd,
which are then converted in 2D ground distances. 
However, it is important to recall that,
due to the pandemic COVID-19, it is forbidden to take new distance measurements 
in the open field.
According to those restrictions and based on our previous research experience, we decided to estimate the 2D ground distances 
(see Fig.~\ref{fig:ranging_precision}) 
between the \ma and the \gd just
calculating the Euclidean distance from the \es to the \home in $(0, 0)$ plus
a random error (overestimation or underestimation) computed as proposed in~\cite{bettisorbelli2018accuracy}\footnote{Our decision is supported by our previous experience in converting slant from/to ground measurements~\cite{bettisorbelli2018accuracy,bettisorbelli2018rangebased}.}.
In other words, we perturb any Euclidean distance $r$ adding the error $E_r$:
\begin{equation}\label{eq:error_ground} 
	E_r = \gamma_d + \frac{h}{r}\gamma_h + e_s \cdot \sqrt{1 + \frac{h^2}{r^2}} = 1.2 + \frac{h}{r} 0.2 + e_s \cdot \sqrt{1 + \frac{h^2}{r^2}} 
\end{equation}
where 
$h$ is the drone's altitude, 
$\gamma_d=1.2 \unit{m}$ is the drone's rolling error, $\gamma_h=0.2$ is the drone's altitude error, and
$e_s$ is a random number in the range $[-\epsilon_s,+\epsilon_s]$,
where $\epsilon_s = 0.1 \unit{m}$ is the UWB instrumental accuracy~\cite{bettisorbelli2018accuracy}.
From Eq.~\eqref{eq:error_ground}, when the ground distance $r=30 \unit{m}$, the maximum ground error $E_r$ is approximately
$1.3 \unit{m}$, $1.4 \unit{m}$, and $1.5 \unit{m}$, for
$h$ is $0 \unit{m}$, $10 \unit{m}$, and $20 \unit{m}$, respectively.
First,  we compare the accuracy of the new RB variant algorithms toward that of their RF implementation. 
Then, we make remarks on the accuracy  of the new RB variants of the three RB algorithms.

\paragraph{Range-free vs Range-based}
In Tab.~\ref{tab:comparison_algorithms_rf_rb} we report the localization accuracy of \xiao, \lee, and \drfe
showing the comparison between the original RF and adapted RB versions of them.
Tab.~\ref{tab:comparison_algorithms_rf_rb} also reports the number of times (in $\%$) 
that an algorithm does not
localize the \gd.
The experimental results take into account the constraints $\rmin = 60 \unit{m}$ and $\amin = 20 \unit{deg}$,
and only VV.
In particular, for the original RF versions we report the average error for both the radii, i.e., 
the observed radius $r=\mu$ and the manufacturer radius $r = r_0$.
For the adapted RB versions we report the average error for the measured radius, denoted as $r = d(A_i, O)$,
where $A_i$ is an \e.

\begin{table}[htbp]
	\caption{Error (in m) and unlocalized (in \%) between RF and RB algorithms in VV.}
	\label{tab:comparison_algorithms_rf_rb}
	\centering
	\begin{tabular}{lc|cc|cc|cc}
     	\multicolumn{2}{c|}{} & \multicolumn{4}{c|}{RF} & \multicolumn{2}{c}{RB} \\
     	\multicolumn{2}{c|}{} & \multicolumn{2}{c|}{$r=\mu$} & \multicolumn{2}{c|}{$r=r_0$} & \multicolumn{2}{c}{$r = d(A_i, O)$} \\
		Algorithm & $h$ & error & unloc & error & unloc & error & unloc \\
		\hline
		\multirow{3}{*}{\xiao} & $0\unit{m}$ & $68$ & $30\%$ & $72$ & $55\%$ & $4$ & $0\%$ \\
		& $10\unit{m}$ & $56$ & $25\%$ & $38$ & $54\%$ & $4$ & $0\%$ \\
		& $20\unit{m}$ & $40$ & $31\%$ & $39$ & $37\%$ & $3$ & $0\%$ \\
		\hline
		\multirow{3}{*}{\lee} & $0\unit{m}$ & $67$ & $30\%$ & $73$ & $55\%$ & $3.8$ & $0\%$ \\
		& $10\unit{m}$ & $55$ & $25\%$ & $38$ & $53\%$ & $3.8$ & $0\%$ \\
		& $20\unit{m}$ & $40$ & $32\%$ & $38$ & $37\%$ & $2.7$ & $0\%$ \\
		\hline
		\multirow{3}{*}{\drfe} & $0\unit{m}$ & $64$ & $29\%$ & $74$ & $55\%$ & $3.2$ & $0\%$ \\
		& $10\unit{m}$ & $55$ & $26\%$ & $37$ & $53\%$ & $3.2$ & $0\%$ \\
		& $20\unit{m}$ & $37$ & $31\%$ & $33$ & $36\%$ & $2.2$ & $0\%$ \\
	\end{tabular}
\end{table}

It is clear that the average error, using measurements, is one order of magnitude smaller.
We have seen that the knowledge of the exact distance significantly improves the accuracy
and practically clear the number of unlocalized devices. 
This also enforces our original statement that the 
knowledge of the radiation pattern is fundamental for the success
of the localization. 
Bubbles and holes in the antenna radiation pattern are no longer a problem because we range the effective distances between the \es and the \gd.
So to avoid measurements, any manufacturer should give as much information as possible on the antenna radiation pattern as possible.

\paragraph{Range-based Comparison}
Here, we present performance-wise the RB versions of the \xiao, \lee, and \drfe algorithms. 
As previously said, \drf is not present because it is not radius-based.
For this simulation evaluation, we consider different cases varying the constraints
of minimum distance $\rmin = \{0, 60\} \unit{m}$ and  
minimum angle $\amin = \{ 20, 40\} \unit{deg}$, 
as explained in Sec.~\ref{sec:simulation}.

\begin{figure}[htbp]
	\centering
	\subfloat[Error: VV, $\rmin = 0 \unit{m}$.]{%
		\centering
		\includegraphics[scale=0.65]{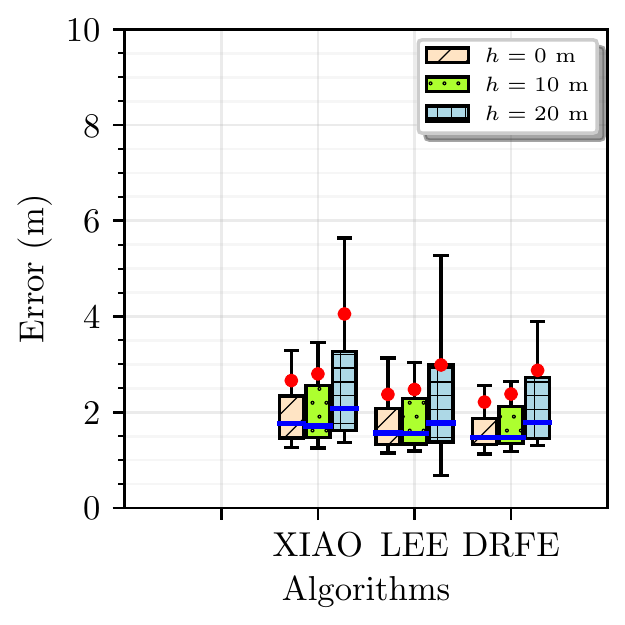}
		\label{fig:error_measurement_VV_4_20_0}
	}
	\subfloat[Error: VH, $\rmin = 0 \unit{m}$.]{%
		\centering
		\includegraphics[scale=0.65]{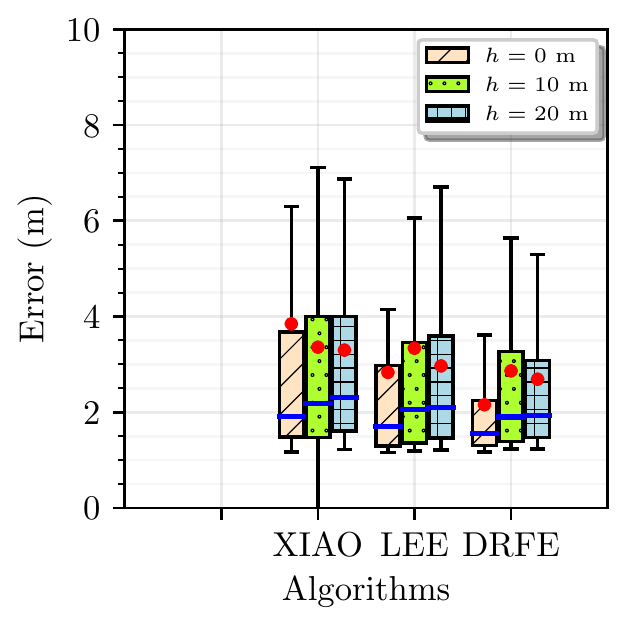}
		\label{fig:error_measurement_VH_4_20_0}
	}
	\subfloat[Error: VV, $\rmin = 60 \unit{m}$.]{%
		\centering
		\includegraphics[scale=0.65]{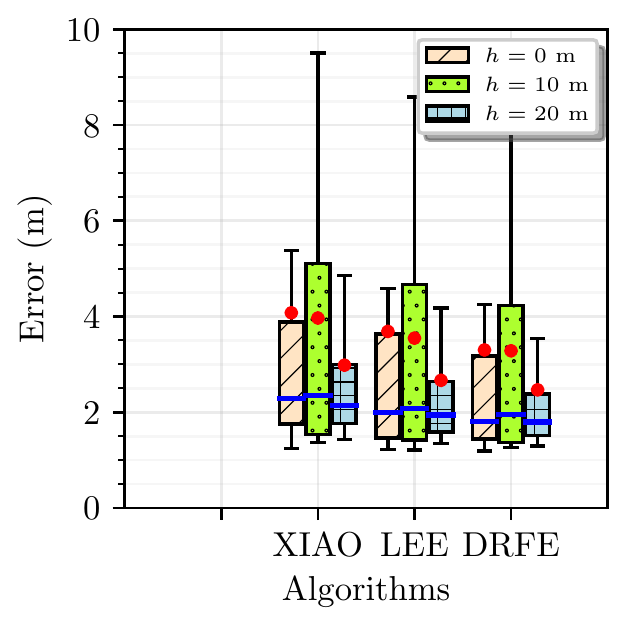}
		\label{fig:error_measurement_VV_4_20_60}
	}
	\subfloat[Error: VH, $\rmin = 60 \unit{m}$.]{%
		\centering
		\includegraphics[scale=0.65]{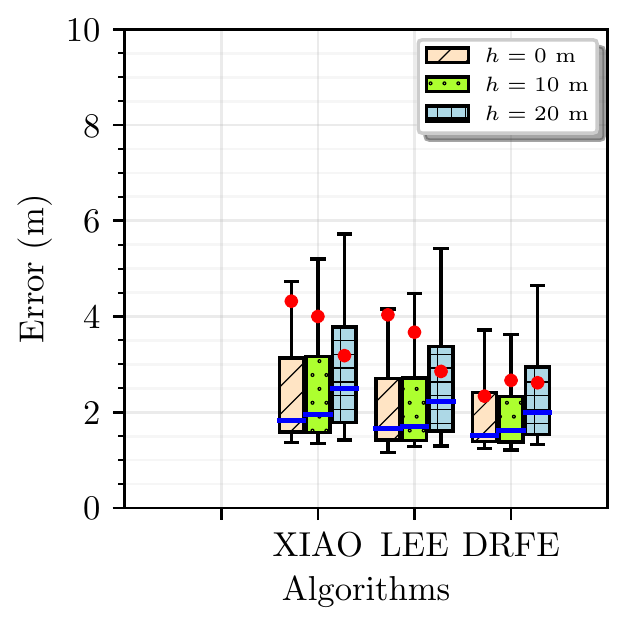}
		\label{fig:error_measurement_VH_4_20_60}
	}
	\caption{Error using distance measurements when $\amin=20 \unit{deg}$ and $\rmin$ varies.}
	\label{fig:error_measurements_20_x}
\end{figure}

\begin{figure}[htbp]
	\centering
	\subfloat[Error: VV, $\rmin = 0 \unit{m}$.]{%
		\centering
		\includegraphics[scale=0.65]{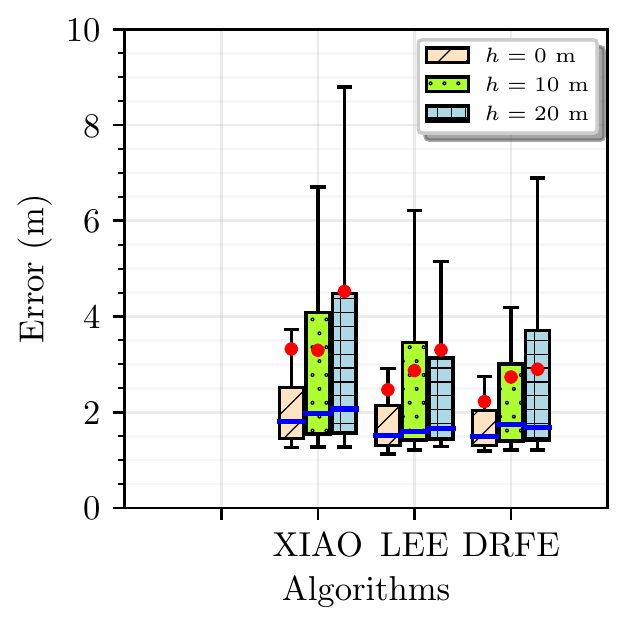}
		\label{fig:error_measurement_VV_4_40_0}
	}
	\subfloat[Error: VH, $\rmin = 0 \unit{m}$.]{%
		\centering
		\includegraphics[scale=0.65]{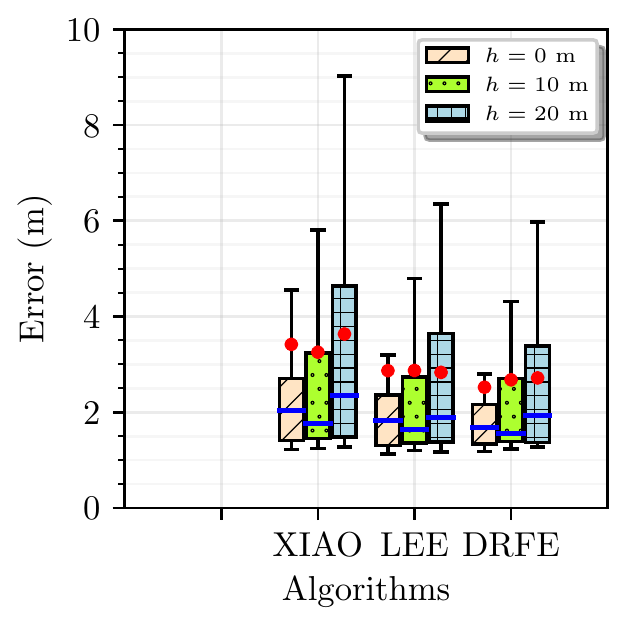}
		\label{fig:error_measurement_VH_4_40_0}
	}
	\subfloat[Error: VV, $\rmin = 60 \unit{m}$.]{%
		\centering
		\includegraphics[scale=0.65]{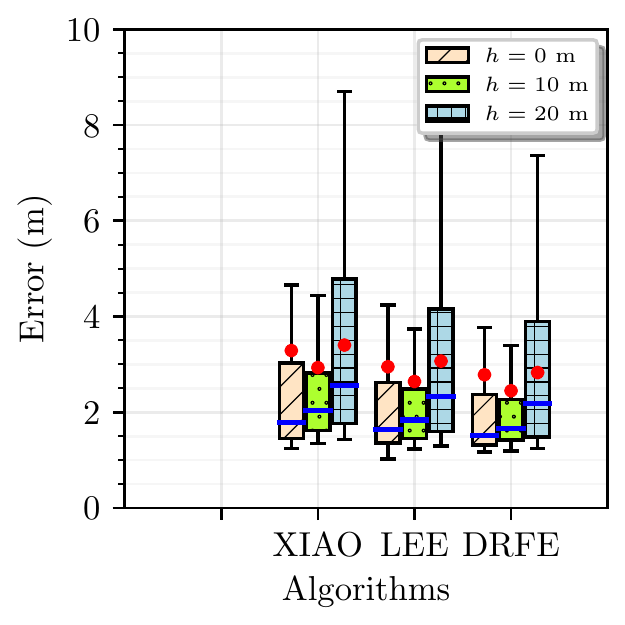}
		\label{fig:error_measurement_VV_4_40_60}
	}
	\subfloat[Error: VH, $\rmin = 60 \unit{m}$.]{%
		\centering
		\includegraphics[scale=0.65]{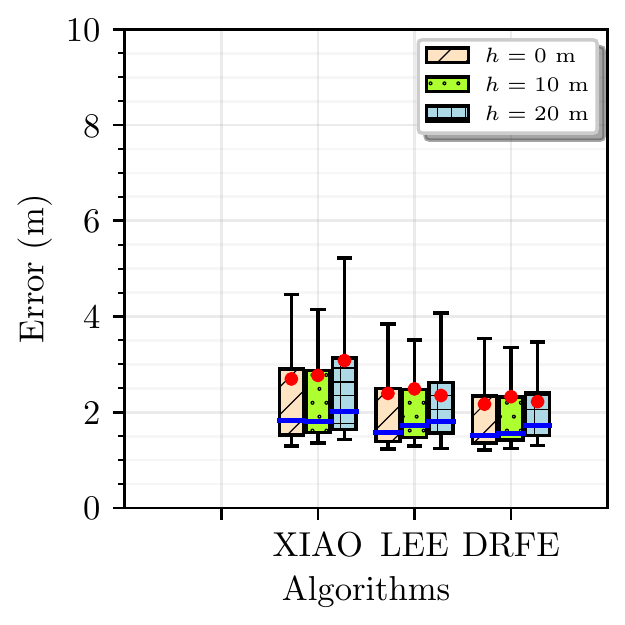}
		\label{fig:error_measurement_VH_4_40_60}
	}
	\caption{Error using distance measurements when $\amin=40 \unit{deg}$ and $\rmin$ varies.}
	\label{fig:error_measurements_40_x}
\end{figure}

In Fig.~\ref{fig:error_measurements_20_x} and Fig.~\ref{fig:error_measurements_40_x}, 
we evaluate the performance of the RB algorithms.
In particular, in Fig.~\ref{fig:error_measurements_20_x} we compare the algorithms
fixing the minimum angle $\amin = 20 \unit{deg}$,
while in Fig.~\ref{fig:error_measurements_40_x} such constraint is kept to $\amin = 40 \unit{deg}$.
Overall, we can see that \drfe works better than the other two algorithms 
with an average error smaller across all the tests, and that \xiao works better than \lee with those settings. 
Also, a better localization accuracy can be obtained for the experiments 
carried on the ground at $h=0\unit{m}$, 
exception made for Fig.~\ref{fig:error_measurement_VV_4_20_60} in which, oddly, 
the error is smaller at $h=20\unit{m}$.
More specifically, in both Fig.~\ref{fig:error_measurements_20_x} and Fig.~\ref{fig:error_measurements_40_x} 
is also evident that VH benefits slightly more from the \rmin constraint of $60 \unit{m}$, 
while it is the opposite for VV which 
behaves better showing smaller error with no \rmin constraint.
Finally, there is very little improvement while testing with larger \amin. 
It is extremely important to note that using measurements, all the selected endpoints are suitable for localizing the \gd in $O$.
This means that the number of unlocalized \gds is zero, 
but also this leads to the growth of the outliers number. 
This can be seen by the presence of long whiskers and a 
considerable difference between the average (i.e., solid circle) and median (i.e., horizontal line) values.
However, even though such whiskers are pretty long, they are really short with respect to the ones
seen in the previous experiments in Sec.~\ref{sec:simulation}, Fig.~\ref{fig:error_real}.

\section{Conclusion}\label{sec:conclusion}
In this paper, we compared the performance of four RF algorithms based on \hnh on a real test-bed using the \dw DWM1001 UWB antennas as \ma and \gds.
We implemented and simulated the algorithms on a large data-set of \es collected in the field.
We analyzed the antenna radiation pattern of the \gd at different altitudes and configurations, i.e., VV and VH.
We have shown how such algorithms actually perform assuming
\begin{inparaenum}[(i)]
	\item first the datasheet radius of the antenna equal to that released by the manufacturer,
	\item then the experimental observed radius, and 
	\item finally the actual radius obtained via distance measurements. 
\end{inparaenum}
The manufacturer datasheet radius poorly performs because usually it does not characterize the antenna radiation pattern very well.
The observed radius can help only if it is almost constant in all the directions.
When the antenna is irregular, only the knowledge of the distances between the \ma and the \gd can alleviate the localization error.
We conclude that the RF algorithms are simple and elegant, but they can be very inaccurate and can leave a high percentage of unlocalized \gds if the antenna is not omnidirectional and measurements are not allowed.
However, the exact knowledge of the irregular antenna can make accurate the RF algorithms.

\bibliographystyle{elsarticle-num}
\bibliography{main-pmc-21} 

\end{document}